\documentclass{article}
\usepackage[utf8]{inputenc}

\usepackage[english]{babel}
\usepackage{amsmath, amssymb, amsthm, amscd,xcolor, graphics, cancel}
\newcommand\ii{{\zeta}}
\newcommand\jj{{\xi}}

\usepackage[pdftex]{graphicx}
\usepackage{hyperref}
\usepackage{cancel}
\usepackage{cite}
\usepackage{alltt}

\usepackage{enumerate}
\newtheorem{theorem}{Theorem}[section]
\newtheorem{definition}[theorem]{Definition}
\newtheorem{example}[theorem]{Example}

\newtheorem{remark}[theorem]{Remark}
\newtheorem{lemma}[theorem]{Lemma}
\newtheorem{corollary}[theorem]{Corollary}
\newtheorem{proposition}[theorem]{Proposition}

\def\fq{\mathbb{F}_{q}}
\def\fqs{\mathbb{F}_{q^2}}

\def\cF{\mathcal{F}}

\def\cF{\mathcal{F}}

\def\cH{\mathcal{H}}
\def\cL{\mathcal{L}}

\def\N{\mathbb{N}}
\def\Z{\mathbb{Z}}

\def\cF{\mathcal{F}}

\newcommand\mut[1]{\ignorespaces}

\def\fq{\mathbb{F}_{q}}
\def\fqs{\mathbb{F}_{q^2}}

\def\cF{\mathcal{F}}

\def\cF{\mathcal{F}}

\def\cH{\mathcal{H}}
\def\cL{\mathcal{L}}

\def\N{\mathbb{N}}
\def\Z{\mathbb{Z}}

\def\cF{\mathcal{F}}
\def\a{\alpha}
\def\b{\beta}
\def\b{{\boldsymbol\beta}}
\def\gg{{\gamma}}
\def\b{\beta}
\def\g{\gamma}
\DeclareMathOperator\supp{supp}

\providecommand\graphkleinD{}
\renewcommand\graphkleinD{
\begin{figure}
\caption{ Klein  function field over ${\mathbb F}_{ 8 }$ defined by the equation $ X^3Y + Y^3Z + XZ^3 =0$. 
Analysis of flags $S_\beta$ satisfying the isometry-dual property.
In this case $n= 21 $, $g= 3 $. We take 
$P=P_\infty=( 1 : 0 : 0 )$, 
$Q_1=( 0 : 1 : 0 )$, 
$Q_2=( 0 : 0 : 1 )$, 
and analyze the codes $C_{{\mathcal L}}(D,aP_\infty+\beta_1Q_1+\beta_2Q_2)$
with $\beta_2= 3 $.
}\label{klein}
\setlength{\unitlength}{1cm}\newcommand\marcaisodual{{\thicklines\circle{1.01}\circle{1.}\circle{.99}}}
\label{fig:QQDKleinisometryduallarge}
\begin{center}
\begin{minipage}{ 1 \textwidth}
\resizebox{\textwidth}{!}{\begin{picture}( 83.2500000000000 , 30.2500000000000 )( -25.1000000000000 , -7.10000000000000 )
\put(0,0){\thicklines\line(1,0){ 36.2600000000000 }}
\put(0,0){\thicklines\line(-1,0){ 22.5750000000000 }}
\put(0,0){\thicklines\line(0,1){ 21.5750000000000 }}
\put(0,0){\thicklines\line(0,-1){ 6.57500000000000 }}
\put( 0 , -1.05000000000000 ){\makebox(0,0){\scalebox{ 3.16250000000000 }{ 0 }}}
\put( 5 , -1.05000000000000 ){\makebox(0,0){\scalebox{ 3.16250000000000 }{ 5 }}}
\put( 10 , -1.05000000000000 ){\makebox(0,0){\scalebox{ 3.16250000000000 }{ 10 }}}
\put( 15 , -1.05000000000000 ){\makebox(0,0){\scalebox{ 3.16250000000000 }{ 15 }}}
\put( 20 , -1.05000000000000 ){\makebox(0,0){\scalebox{ 3.16250000000000 }{ 20 }}}
\put( 25 , -1.05000000000000 ){\makebox(0,0){\scalebox{ 3.16250000000000 }{ 25 }}}
\put( 30 , -1.05000000000000 ){\makebox(0,0){\scalebox{ 3.16250000000000 }{ 30 }}}
\put( 35 , -1.05000000000000 ){\makebox(0,0){\scalebox{ 3.16250000000000 }{ 35 }}}
\put( -5 , -1.05000000000000 ){\makebox(0,0){\scalebox{ 3.16250000000000 }{ -5 }}}
\put( -10 , -1.05000000000000 ){\makebox(0,0){\scalebox{ 3.16250000000000 }{ -10 }}}
\put( -15 , -1.05000000000000 ){\makebox(0,0){\scalebox{ 3.16250000000000 }{ -15 }}}
\put( -20 , -1.05000000000000 ){\makebox(0,0){\scalebox{ 3.16250000000000 }{ -20 }}}
\put( -1.05000000000000 , 0 ){\makebox(0,0){\scalebox{ 3.16250000000000 }{ 0 }}}
\put( -1.05000000000000 , 5 ){\makebox(0,0){\scalebox{ 3.16250000000000 }{ 5 }}}
\put( -1.05000000000000 , 10 ){\makebox(0,0){\scalebox{ 3.16250000000000 }{ 10 }}}
\put( -1.05000000000000 , 15 ){\makebox(0,0){\scalebox{ 3.16250000000000 }{ 15 }}}
\put( -1.05000000000000 , 20 ){\makebox(0,0){\scalebox{ 3.16250000000000 }{ 20 }}}
\put( -1.05000000000000 , -5 ){\makebox(0,0){\scalebox{ 3.16250000000000 }{ -5 }}}
\put( -0.262500000000000 , 22.1000000000000 ){{\scalebox{ 4.74375000000000 }{$\beta_1$}}}
\put( 37.1000000000000 , -0.262500000000000 ){{\scalebox{ 4.74375000000000 }{$a$}}}
\put( 0 , 21.5750000000000 ){\makebox(0,0){\scalebox{ 3.16250000000000 }{$\blacktriangle$}}}
\put( 36.3125000000000 , 0 ){\makebox(0,0){\scalebox{ 3.16250000000000 }{$\blacktriangleright$}}}
\put( 4 , -5 ){{\makebox(0,0){\scalebox{3}{$*$}}}}
\put( 6 , -5 ){{\makebox(0,0){\scalebox{3}{$*$}}}}
\put( 7 , -5 ){{\makebox(0,0){\scalebox{3}{$*$}}}}
\put( 8 , -5 ){{\makebox(0,0){\scalebox{3}{$*$}}}}
\put( 9 , -5 ){{\makebox(0,0){\scalebox{3}{$*$}}}}
\put( 10 , -5 ){{\makebox(0,0){\scalebox{3}{$*$}}}}
\put( 11 , -5 ){{\makebox(0,0){\scalebox{3}{$*$}}}}
\put( 12 , -5 ){{\makebox(0,0){\scalebox{3}{$*$}}}}
\put( 13 , -5 ){{\makebox(0,0){\scalebox{3}{$*$}}}}
\put( 14 , -5 ){{\makebox(0,0){\scalebox{3}{$*$}}}}
\put( 15 , -5 ){{\makebox(0,0){\scalebox{3}{$*$}}}}
\put( 16 , -5 ){{\makebox(0,0){\scalebox{3}{$*$}}}}
\put( 17 , -5 ){{\makebox(0,0){\scalebox{3}{$*$}}}}
\put( 18 , -5 ){{\makebox(0,0){\scalebox{3}{$*$}}}}
\put( 19 , -5 ){{\makebox(0,0){\scalebox{3}{$*$}}}}
\put( 20 , -5 ){{\makebox(0,0){\scalebox{3}{$*$}}}}
\put( 21 , -5 ){{\makebox(0,0){\scalebox{3}{$*$}}}}
\put( 22 , -5 ){{\makebox(0,0){\scalebox{3}{$*$}}}}
\put( 23 , -5 ){{\makebox(0,0){\scalebox{3}{$*$}}}}
\put( 24 , -5 ){{\makebox(0,0){\scalebox{3}{$*$}}}}
\put( 26 , -5 ){{\makebox(0,0){\scalebox{3}{$*$}}}}
\put( 4 , -4 ){{\makebox(0,0){\scalebox{3}{$*$}}}}
\put( 5 , -4 ){{\makebox(0,0){\scalebox{3}{$*$}}}}
\put( 6 , -4 ){{\makebox(0,0){\scalebox{3}{$*$}}}}
\put( 7 , -4 ){{\makebox(0,0){\scalebox{3}{$*$}}}}
\put( 8 , -4 ){{\makebox(0,0){\scalebox{3}{$*$}}}}
\put( 9 , -4 ){{\makebox(0,0){\scalebox{3}{$*$}}}}
\put( 10 , -4 ){{\makebox(0,0){\scalebox{3}{$*$}}}}
\put( 11 , -4 ){{\makebox(0,0){\scalebox{3}{$*$}}}}
\put( 12 , -4 ){{\makebox(0,0){\scalebox{3}{$*$}}}}
\put( 13 , -4 ){{\makebox(0,0){\scalebox{3}{$*$}}}}
\put( 14 , -4 ){{\makebox(0,0){\scalebox{3}{$*$}}}}
\put( 15 , -4 ){{\makebox(0,0){\scalebox{3}{$*$}}}}
\put( 16 , -4 ){{\makebox(0,0){\scalebox{3}{$*$}}}}
\put( 17 , -4 ){{\makebox(0,0){\scalebox{3}{$*$}}}}
\put( 18 , -4 ){{\makebox(0,0){\scalebox{3}{$*$}}}}
\put( 19 , -4 ){{\makebox(0,0){\scalebox{3}{$*$}}}}
\put( 20 , -4 ){{\makebox(0,0){\scalebox{3}{$*$}}}}
\put( 21 , -4 ){{\makebox(0,0){\scalebox{3}{$*$}}}}
\put( 22 , -4 ){{\makebox(0,0){\scalebox{3}{$*$}}}}
\put( 23 , -4 ){{\makebox(0,0){\scalebox{3}{$*$}}}}
\put( 24 , -4 ){{\makebox(0,0){\scalebox{3}{$*$}}}}
\put( 4 , -4 ){\circle*{4.5}}
\put( 5 , -4 ){\circle*{4.5}}
\put( 6 , -4 ){\circle*{4.5}}
\put( 7 , -4 ){\circle*{4.5}}
\put( 8 , -4 ){\circle*{4.5}}
\put( 9 , -4 ){\circle*{4.5}}
\put( 10 , -4 ){\circle*{4.5}}
\put( 11 , -4 ){\circle*{4.5}}
\put( 12 , -4 ){\circle*{4.5}}
\put( 13 , -4 ){\circle*{4.5}}
\put( 14 , -4 ){\circle*{4.5}}
\put( 15 , -4 ){\circle*{4.5}}
\put( 16 , -4 ){\circle*{4.5}}
\put( 17 , -4 ){\circle*{4.5}}
\put( 18 , -4 ){\circle*{4.5}}
\put( 19 , -4 ){\circle*{4.5}}
\put( 20 , -4 ){\circle*{4.5}}
\put( 21 , -4 ){\circle*{4.5}}
\put( 22 , -4 ){\circle*{4.5}}
\put( 23 , -4 ){\circle*{4.5}}
\put( 25 , -4 ){\scalebox{4}{${\bf x}_{( -4 , 3 )}= (1,\alpha^{3},\alpha^{2},\alpha^{3},\alpha^{2},1,1,\alpha^{2},\alpha^{3},\alpha^{2},1,\alpha^{3},\alpha^{3},1,\alpha^{2},\alpha^{2},\alpha^{3},1,\alpha^{2},1,\alpha^{3}) $}}
\put( 2 , -3 ){{\makebox(0,0){\scalebox{3}{$*$}}}}
\put( 4 , -3 ){{\makebox(0,0){\scalebox{3}{$*$}}}}
\put( 5 , -3 ){{\makebox(0,0){\scalebox{3}{$*$}}}}
\put( 6 , -3 ){{\makebox(0,0){\scalebox{3}{$*$}}}}
\put( 7 , -3 ){{\makebox(0,0){\scalebox{3}{$*$}}}}
\put( 8 , -3 ){{\makebox(0,0){\scalebox{3}{$*$}}}}
\put( 9 , -3 ){{\makebox(0,0){\scalebox{3}{$*$}}}}
\put( 10 , -3 ){{\makebox(0,0){\scalebox{3}{$*$}}}}
\put( 11 , -3 ){{\makebox(0,0){\scalebox{3}{$*$}}}}
\put( 12 , -3 ){{\makebox(0,0){\scalebox{3}{$*$}}}}
\put( 13 , -3 ){{\makebox(0,0){\scalebox{3}{$*$}}}}
\put( 14 , -3 ){{\makebox(0,0){\scalebox{3}{$*$}}}}
\put( 15 , -3 ){{\makebox(0,0){\scalebox{3}{$*$}}}}
\put( 16 , -3 ){{\makebox(0,0){\scalebox{3}{$*$}}}}
\put( 17 , -3 ){{\makebox(0,0){\scalebox{3}{$*$}}}}
\put( 18 , -3 ){{\makebox(0,0){\scalebox{3}{$*$}}}}
\put( 19 , -3 ){{\makebox(0,0){\scalebox{3}{$*$}}}}
\put( 20 , -3 ){{\makebox(0,0){\scalebox{3}{$*$}}}}
\put( 21 , -3 ){{\makebox(0,0){\scalebox{3}{$*$}}}}
\put( 22 , -3 ){{\makebox(0,0){\scalebox{3}{$*$}}}}
\put( 24 , -3 ){{\makebox(0,0){\scalebox{3}{$*$}}}}
\put( -1 , -2 ){{\makebox(0,0){\scalebox{3}{$*$}}}}
\put( 2 , -2 ){{\makebox(0,0){\scalebox{3}{$*$}}}}
\put( 4 , -2 ){{\makebox(0,0){\scalebox{3}{$*$}}}}
\put( 5 , -2 ){{\makebox(0,0){\scalebox{3}{$*$}}}}
\put( 6 , -2 ){{\makebox(0,0){\scalebox{3}{$*$}}}}
\put( 7 , -2 ){{\makebox(0,0){\scalebox{3}{$*$}}}}
\put( 8 , -2 ){{\makebox(0,0){\scalebox{3}{$*$}}}}
\put( 9 , -2 ){{\makebox(0,0){\scalebox{3}{$*$}}}}
\put( 10 , -2 ){{\makebox(0,0){\scalebox{3}{$*$}}}}
\put( 11 , -2 ){{\makebox(0,0){\scalebox{3}{$*$}}}}
\put( 12 , -2 ){{\makebox(0,0){\scalebox{3}{$*$}}}}
\put( 13 , -2 ){{\makebox(0,0){\scalebox{3}{$*$}}}}
\put( 14 , -2 ){{\makebox(0,0){\scalebox{3}{$*$}}}}
\put( 15 , -2 ){{\makebox(0,0){\scalebox{3}{$*$}}}}
\put( 16 , -2 ){{\makebox(0,0){\scalebox{3}{$*$}}}}
\put( 17 , -2 ){{\makebox(0,0){\scalebox{3}{$*$}}}}
\put( 18 , -2 ){{\makebox(0,0){\scalebox{3}{$*$}}}}
\put( 19 , -2 ){{\makebox(0,0){\scalebox{3}{$*$}}}}
\put( 21 , -2 ){{\makebox(0,0){\scalebox{3}{$*$}}}}
\put( 22 , -2 ){{\makebox(0,0){\scalebox{3}{$*$}}}}
\put( 24 , -2 ){{\makebox(0,0){\scalebox{3}{$*$}}}}
\put( -1 , -1 ){{\makebox(0,0){\scalebox{3}{$*$}}}}
\put( 2 , -1 ){{\makebox(0,0){\scalebox{3}{$*$}}}}
\put( 3 , -1 ){{\makebox(0,0){\scalebox{3}{$*$}}}}
\put( 4 , -1 ){{\makebox(0,0){\scalebox{3}{$*$}}}}
\put( 5 , -1 ){{\makebox(0,0){\scalebox{3}{$*$}}}}
\put( 6 , -1 ){{\makebox(0,0){\scalebox{3}{$*$}}}}
\put( 7 , -1 ){{\makebox(0,0){\scalebox{3}{$*$}}}}
\put( 8 , -1 ){{\makebox(0,0){\scalebox{3}{$*$}}}}
\put( 9 , -1 ){{\makebox(0,0){\scalebox{3}{$*$}}}}
\put( 10 , -1 ){{\makebox(0,0){\scalebox{3}{$*$}}}}
\put( 11 , -1 ){{\makebox(0,0){\scalebox{3}{$*$}}}}
\put( 12 , -1 ){{\makebox(0,0){\scalebox{3}{$*$}}}}
\put( 13 , -1 ){{\makebox(0,0){\scalebox{3}{$*$}}}}
\put( 14 , -1 ){{\makebox(0,0){\scalebox{3}{$*$}}}}
\put( 15 , -1 ){{\makebox(0,0){\scalebox{3}{$*$}}}}
\put( 16 , -1 ){{\makebox(0,0){\scalebox{3}{$*$}}}}
\put( 17 , -1 ){{\makebox(0,0){\scalebox{3}{$*$}}}}
\put( 18 , -1 ){{\makebox(0,0){\scalebox{3}{$*$}}}}
\put( 19 , -1 ){{\makebox(0,0){\scalebox{3}{$*$}}}}
\put( 21 , -1 ){{\makebox(0,0){\scalebox{3}{$*$}}}}
\put( 22 , -1 ){{\makebox(0,0){\scalebox{3}{$*$}}}}
\put( -1 , 0 ){{\makebox(0,0){\scalebox{3}{$*$}}}}
\put( 0 , 0 ){{\makebox(0,0){\scalebox{3}{$*$}}}}
\put( 2 , 0 ){{\makebox(0,0){\scalebox{3}{$*$}}}}
\put( 3 , 0 ){{\makebox(0,0){\scalebox{3}{$*$}}}}
\put( 4 , 0 ){{\makebox(0,0){\scalebox{3}{$*$}}}}
\put( 5 , 0 ){{\makebox(0,0){\scalebox{3}{$*$}}}}
\put( 6 , 0 ){{\makebox(0,0){\scalebox{3}{$*$}}}}
\put( 7 , 0 ){{\makebox(0,0){\scalebox{3}{$*$}}}}
\put( 8 , 0 ){{\makebox(0,0){\scalebox{3}{$*$}}}}
\put( 9 , 0 ){{\makebox(0,0){\scalebox{3}{$*$}}}}
\put( 10 , 0 ){{\makebox(0,0){\scalebox{3}{$*$}}}}
\put( 11 , 0 ){{\makebox(0,0){\scalebox{3}{$*$}}}}
\put( 12 , 0 ){{\makebox(0,0){\scalebox{3}{$*$}}}}
\put( 13 , 0 ){{\makebox(0,0){\scalebox{3}{$*$}}}}
\put( 14 , 0 ){{\makebox(0,0){\scalebox{3}{$*$}}}}
\put( 15 , 0 ){{\makebox(0,0){\scalebox{3}{$*$}}}}
\put( 16 , 0 ){{\makebox(0,0){\scalebox{3}{$*$}}}}
\put( 17 , 0 ){{\makebox(0,0){\scalebox{3}{$*$}}}}
\put( 18 , 0 ){{\makebox(0,0){\scalebox{3}{$*$}}}}
\put( 19 , 0 ){{\makebox(0,0){\scalebox{3}{$*$}}}}
\put( 22 , 0 ){{\makebox(0,0){\scalebox{3}{$*$}}}}
\put( -3 , 1 ){{\makebox(0,0){\scalebox{3}{$*$}}}}
\put( -1 , 1 ){{\makebox(0,0){\scalebox{3}{$*$}}}}
\put( 0 , 1 ){{\makebox(0,0){\scalebox{3}{$*$}}}}
\put( 2 , 1 ){{\makebox(0,0){\scalebox{3}{$*$}}}}
\put( 3 , 1 ){{\makebox(0,0){\scalebox{3}{$*$}}}}
\put( 4 , 1 ){{\makebox(0,0){\scalebox{3}{$*$}}}}
\put( 5 , 1 ){{\makebox(0,0){\scalebox{3}{$*$}}}}
\put( 6 , 1 ){{\makebox(0,0){\scalebox{3}{$*$}}}}
\put( 7 , 1 ){{\makebox(0,0){\scalebox{3}{$*$}}}}
\put( 8 , 1 ){{\makebox(0,0){\scalebox{3}{$*$}}}}
\put( 9 , 1 ){{\makebox(0,0){\scalebox{3}{$*$}}}}
\put( 10 , 1 ){{\makebox(0,0){\scalebox{3}{$*$}}}}
\put( 11 , 1 ){{\makebox(0,0){\scalebox{3}{$*$}}}}
\put( 12 , 1 ){{\makebox(0,0){\scalebox{3}{$*$}}}}
\put( 13 , 1 ){{\makebox(0,0){\scalebox{3}{$*$}}}}
\put( 14 , 1 ){{\makebox(0,0){\scalebox{3}{$*$}}}}
\put( 15 , 1 ){{\makebox(0,0){\scalebox{3}{$*$}}}}
\put( 16 , 1 ){{\makebox(0,0){\scalebox{3}{$*$}}}}
\put( 17 , 1 ){{\makebox(0,0){\scalebox{3}{$*$}}}}
\put( 19 , 1 ){{\makebox(0,0){\scalebox{3}{$*$}}}}
\put( 22 , 1 ){{\makebox(0,0){\scalebox{3}{$*$}}}}
\put( -3 , 2 ){{\makebox(0,0){\scalebox{3}{$*$}}}}
\put( -1 , 2 ){{\makebox(0,0){\scalebox{3}{$*$}}}}
\put( 0 , 2 ){{\makebox(0,0){\scalebox{3}{$*$}}}}
\put( 1 , 2 ){{\makebox(0,0){\scalebox{3}{$*$}}}}
\put( 2 , 2 ){{\makebox(0,0){\scalebox{3}{$*$}}}}
\put( 3 , 2 ){{\makebox(0,0){\scalebox{3}{$*$}}}}
\put( 4 , 2 ){{\makebox(0,0){\scalebox{3}{$*$}}}}
\put( 5 , 2 ){{\makebox(0,0){\scalebox{3}{$*$}}}}
\put( 6 , 2 ){{\makebox(0,0){\scalebox{3}{$*$}}}}
\put( 7 , 2 ){{\makebox(0,0){\scalebox{3}{$*$}}}}
\put( 8 , 2 ){{\makebox(0,0){\scalebox{3}{$*$}}}}
\put( 9 , 2 ){{\makebox(0,0){\scalebox{3}{$*$}}}}
\put( 10 , 2 ){{\makebox(0,0){\scalebox{3}{$*$}}}}
\put( 11 , 2 ){{\makebox(0,0){\scalebox{3}{$*$}}}}
\put( 12 , 2 ){{\makebox(0,0){\scalebox{3}{$*$}}}}
\put( 13 , 2 ){{\makebox(0,0){\scalebox{3}{$*$}}}}
\put( 14 , 2 ){{\makebox(0,0){\scalebox{3}{$*$}}}}
\put( 15 , 2 ){{\makebox(0,0){\scalebox{3}{$*$}}}}
\put( 16 , 2 ){{\makebox(0,0){\scalebox{3}{$*$}}}}
\put( 17 , 2 ){{\makebox(0,0){\scalebox{3}{$*$}}}}
\put( 19 , 2 ){{\makebox(0,0){\scalebox{3}{$*$}}}}
\put( -3 , 3 ){{\makebox(0,0){\scalebox{3}{$*$}}}}
\put( -2 , 3 ){{\makebox(0,0){\scalebox{3}{$*$}}}}
\put( -1 , 3 ){{\makebox(0,0){\scalebox{3}{$*$}}}}
\put( 0 , 3 ){{\makebox(0,0){\scalebox{3}{$*$}}}}
\put( 1 , 3 ){{\makebox(0,0){\scalebox{3}{$*$}}}}
\put( 2 , 3 ){{\makebox(0,0){\scalebox{3}{$*$}}}}
\put( 3 , 3 ){{\makebox(0,0){\scalebox{3}{$*$}}}}
\put( 4 , 3 ){{\makebox(0,0){\scalebox{3}{$*$}}}}
\put( 5 , 3 ){{\makebox(0,0){\scalebox{3}{$*$}}}}
\put( 6 , 3 ){{\makebox(0,0){\scalebox{3}{$*$}}}}
\put( 7 , 3 ){{\makebox(0,0){\scalebox{3}{$*$}}}}
\put( 8 , 3 ){{\makebox(0,0){\scalebox{3}{$*$}}}}
\put( 9 , 3 ){{\makebox(0,0){\scalebox{3}{$*$}}}}
\put( 10 , 3 ){{\makebox(0,0){\scalebox{3}{$*$}}}}
\put( 11 , 3 ){{\makebox(0,0){\scalebox{3}{$*$}}}}
\put( 12 , 3 ){{\makebox(0,0){\scalebox{3}{$*$}}}}
\put( 13 , 3 ){{\makebox(0,0){\scalebox{3}{$*$}}}}
\put( 14 , 3 ){{\makebox(0,0){\scalebox{3}{$*$}}}}
\put( 15 , 3 ){{\makebox(0,0){\scalebox{3}{$*$}}}}
\put( 16 , 3 ){{\makebox(0,0){\scalebox{3}{$*$}}}}
\put( 17 , 3 ){{\makebox(0,0){\scalebox{3}{$*$}}}}
\put( -3 , 3 ){\circle*{4.5}}
\put( -2 , 3 ){\circle*{4.5}}
\put( -1 , 3 ){\circle*{4.5}}
\put( 0 , 3 ){\circle*{4.5}}
\put( 1 , 3 ){\circle*{4.5}}
\put( 2 , 3 ){\circle*{4.5}}
\put( 3 , 3 ){\circle*{4.5}}
\put( 4 , 3 ){\circle*{4.5}}
\put( 5 , 3 ){\circle*{4.5}}
\put( 6 , 3 ){\circle*{4.5}}
\put( 7 , 3 ){\circle*{4.5}}
\put( 8 , 3 ){\circle*{4.5}}
\put( 9 , 3 ){\circle*{4.5}}
\put( 10 , 3 ){\circle*{4.5}}
\put( 11 , 3 ){\circle*{4.5}}
\put( 12 , 3 ){\circle*{4.5}}
\put( 13 , 3 ){\circle*{4.5}}
\put( 14 , 3 ){\circle*{4.5}}
\put( 15 , 3 ){\circle*{4.5}}
\put( 16 , 3 ){\circle*{4.5}}
\put( 18 , 3 ){\scalebox{4}{${\bf x}_{( 3 , 3 )}= (1,\alpha^{4},\alpha^{5},\alpha^{4},\alpha^{5},1,1,\alpha^{5},\alpha^{4},\alpha^{5},1,\alpha^{4},\alpha^{4},1,\alpha^{5},\alpha^{5},\alpha^{4},1,\alpha^{5},1,\alpha^{4}) $}}
\put( -5 , 4 ){{\makebox(0,0){\scalebox{3}{$*$}}}}
\put( -3 , 4 ){{\makebox(0,0){\scalebox{3}{$*$}}}}
\put( -2 , 4 ){{\makebox(0,0){\scalebox{3}{$*$}}}}
\put( -1 , 4 ){{\makebox(0,0){\scalebox{3}{$*$}}}}
\put( 0 , 4 ){{\makebox(0,0){\scalebox{3}{$*$}}}}
\put( 1 , 4 ){{\makebox(0,0){\scalebox{3}{$*$}}}}
\put( 2 , 4 ){{\makebox(0,0){\scalebox{3}{$*$}}}}
\put( 3 , 4 ){{\makebox(0,0){\scalebox{3}{$*$}}}}
\put( 4 , 4 ){{\makebox(0,0){\scalebox{3}{$*$}}}}
\put( 5 , 4 ){{\makebox(0,0){\scalebox{3}{$*$}}}}
\put( 6 , 4 ){{\makebox(0,0){\scalebox{3}{$*$}}}}
\put( 7 , 4 ){{\makebox(0,0){\scalebox{3}{$*$}}}}
\put( 8 , 4 ){{\makebox(0,0){\scalebox{3}{$*$}}}}
\put( 9 , 4 ){{\makebox(0,0){\scalebox{3}{$*$}}}}
\put( 10 , 4 ){{\makebox(0,0){\scalebox{3}{$*$}}}}
\put( 11 , 4 ){{\makebox(0,0){\scalebox{3}{$*$}}}}
\put( 12 , 4 ){{\makebox(0,0){\scalebox{3}{$*$}}}}
\put( 13 , 4 ){{\makebox(0,0){\scalebox{3}{$*$}}}}
\put( 14 , 4 ){{\makebox(0,0){\scalebox{3}{$*$}}}}
\put( 15 , 4 ){{\makebox(0,0){\scalebox{3}{$*$}}}}
\put( 17 , 4 ){{\makebox(0,0){\scalebox{3}{$*$}}}}
\put( -8 , 5 ){{\makebox(0,0){\scalebox{3}{$*$}}}}
\put( -5 , 5 ){{\makebox(0,0){\scalebox{3}{$*$}}}}
\put( -3 , 5 ){{\makebox(0,0){\scalebox{3}{$*$}}}}
\put( -2 , 5 ){{\makebox(0,0){\scalebox{3}{$*$}}}}
\put( -1 , 5 ){{\makebox(0,0){\scalebox{3}{$*$}}}}
\put( 0 , 5 ){{\makebox(0,0){\scalebox{3}{$*$}}}}
\put( 1 , 5 ){{\makebox(0,0){\scalebox{3}{$*$}}}}
\put( 2 , 5 ){{\makebox(0,0){\scalebox{3}{$*$}}}}
\put( 3 , 5 ){{\makebox(0,0){\scalebox{3}{$*$}}}}
\put( 4 , 5 ){{\makebox(0,0){\scalebox{3}{$*$}}}}
\put( 5 , 5 ){{\makebox(0,0){\scalebox{3}{$*$}}}}
\put( 6 , 5 ){{\makebox(0,0){\scalebox{3}{$*$}}}}
\put( 7 , 5 ){{\makebox(0,0){\scalebox{3}{$*$}}}}
\put( 8 , 5 ){{\makebox(0,0){\scalebox{3}{$*$}}}}
\put( 9 , 5 ){{\makebox(0,0){\scalebox{3}{$*$}}}}
\put( 10 , 5 ){{\makebox(0,0){\scalebox{3}{$*$}}}}
\put( 11 , 5 ){{\makebox(0,0){\scalebox{3}{$*$}}}}
\put( 12 , 5 ){{\makebox(0,0){\scalebox{3}{$*$}}}}
\put( 14 , 5 ){{\makebox(0,0){\scalebox{3}{$*$}}}}
\put( 15 , 5 ){{\makebox(0,0){\scalebox{3}{$*$}}}}
\put( 17 , 5 ){{\makebox(0,0){\scalebox{3}{$*$}}}}
\put( -8 , 6 ){{\makebox(0,0){\scalebox{3}{$*$}}}}
\put( -5 , 6 ){{\makebox(0,0){\scalebox{3}{$*$}}}}
\put( -4 , 6 ){{\makebox(0,0){\scalebox{3}{$*$}}}}
\put( -3 , 6 ){{\makebox(0,0){\scalebox{3}{$*$}}}}
\put( -2 , 6 ){{\makebox(0,0){\scalebox{3}{$*$}}}}
\put( -1 , 6 ){{\makebox(0,0){\scalebox{3}{$*$}}}}
\put( 0 , 6 ){{\makebox(0,0){\scalebox{3}{$*$}}}}
\put( 1 , 6 ){{\makebox(0,0){\scalebox{3}{$*$}}}}
\put( 2 , 6 ){{\makebox(0,0){\scalebox{3}{$*$}}}}
\put( 3 , 6 ){{\makebox(0,0){\scalebox{3}{$*$}}}}
\put( 4 , 6 ){{\makebox(0,0){\scalebox{3}{$*$}}}}
\put( 5 , 6 ){{\makebox(0,0){\scalebox{3}{$*$}}}}
\put( 6 , 6 ){{\makebox(0,0){\scalebox{3}{$*$}}}}
\put( 7 , 6 ){{\makebox(0,0){\scalebox{3}{$*$}}}}
\put( 8 , 6 ){{\makebox(0,0){\scalebox{3}{$*$}}}}
\put( 9 , 6 ){{\makebox(0,0){\scalebox{3}{$*$}}}}
\put( 10 , 6 ){{\makebox(0,0){\scalebox{3}{$*$}}}}
\put( 11 , 6 ){{\makebox(0,0){\scalebox{3}{$*$}}}}
\put( 12 , 6 ){{\makebox(0,0){\scalebox{3}{$*$}}}}
\put( 14 , 6 ){{\makebox(0,0){\scalebox{3}{$*$}}}}
\put( 15 , 6 ){{\makebox(0,0){\scalebox{3}{$*$}}}}
\put( -8 , 7 ){{\makebox(0,0){\scalebox{3}{$*$}}}}
\put( -7 , 7 ){{\makebox(0,0){\scalebox{3}{$*$}}}}
\put( -5 , 7 ){{\makebox(0,0){\scalebox{3}{$*$}}}}
\put( -4 , 7 ){{\makebox(0,0){\scalebox{3}{$*$}}}}
\put( -3 , 7 ){{\makebox(0,0){\scalebox{3}{$*$}}}}
\put( -2 , 7 ){{\makebox(0,0){\scalebox{3}{$*$}}}}
\put( -1 , 7 ){{\makebox(0,0){\scalebox{3}{$*$}}}}
\put( 0 , 7 ){{\makebox(0,0){\scalebox{3}{$*$}}}}
\put( 1 , 7 ){{\makebox(0,0){\scalebox{3}{$*$}}}}
\put( 2 , 7 ){{\makebox(0,0){\scalebox{3}{$*$}}}}
\put( 3 , 7 ){{\makebox(0,0){\scalebox{3}{$*$}}}}
\put( 4 , 7 ){{\makebox(0,0){\scalebox{3}{$*$}}}}
\put( 5 , 7 ){{\makebox(0,0){\scalebox{3}{$*$}}}}
\put( 6 , 7 ){{\makebox(0,0){\scalebox{3}{$*$}}}}
\put( 7 , 7 ){{\makebox(0,0){\scalebox{3}{$*$}}}}
\put( 8 , 7 ){{\makebox(0,0){\scalebox{3}{$*$}}}}
\put( 9 , 7 ){{\makebox(0,0){\scalebox{3}{$*$}}}}
\put( 10 , 7 ){{\makebox(0,0){\scalebox{3}{$*$}}}}
\put( 11 , 7 ){{\makebox(0,0){\scalebox{3}{$*$}}}}
\put( 12 , 7 ){{\makebox(0,0){\scalebox{3}{$*$}}}}
\put( 15 , 7 ){{\makebox(0,0){\scalebox{3}{$*$}}}}
\put( -10 , 8 ){{\makebox(0,0){\scalebox{3}{$*$}}}}
\put( -8 , 8 ){{\makebox(0,0){\scalebox{3}{$*$}}}}
\put( -7 , 8 ){{\makebox(0,0){\scalebox{3}{$*$}}}}
\put( -5 , 8 ){{\makebox(0,0){\scalebox{3}{$*$}}}}
\put( -4 , 8 ){{\makebox(0,0){\scalebox{3}{$*$}}}}
\put( -3 , 8 ){{\makebox(0,0){\scalebox{3}{$*$}}}}
\put( -2 , 8 ){{\makebox(0,0){\scalebox{3}{$*$}}}}
\put( -1 , 8 ){{\makebox(0,0){\scalebox{3}{$*$}}}}
\put( 0 , 8 ){{\makebox(0,0){\scalebox{3}{$*$}}}}
\put( 1 , 8 ){{\makebox(0,0){\scalebox{3}{$*$}}}}
\put( 2 , 8 ){{\makebox(0,0){\scalebox{3}{$*$}}}}
\put( 3 , 8 ){{\makebox(0,0){\scalebox{3}{$*$}}}}
\put( 4 , 8 ){{\makebox(0,0){\scalebox{3}{$*$}}}}
\put( 5 , 8 ){{\makebox(0,0){\scalebox{3}{$*$}}}}
\put( 6 , 8 ){{\makebox(0,0){\scalebox{3}{$*$}}}}
\put( 7 , 8 ){{\makebox(0,0){\scalebox{3}{$*$}}}}
\put( 8 , 8 ){{\makebox(0,0){\scalebox{3}{$*$}}}}
\put( 9 , 8 ){{\makebox(0,0){\scalebox{3}{$*$}}}}
\put( 10 , 8 ){{\makebox(0,0){\scalebox{3}{$*$}}}}
\put( 12 , 8 ){{\makebox(0,0){\scalebox{3}{$*$}}}}
\put( 15 , 8 ){{\makebox(0,0){\scalebox{3}{$*$}}}}
\put( -10 , 9 ){{\makebox(0,0){\scalebox{3}{$*$}}}}
\put( -8 , 9 ){{\makebox(0,0){\scalebox{3}{$*$}}}}
\put( -7 , 9 ){{\makebox(0,0){\scalebox{3}{$*$}}}}
\put( -6 , 9 ){{\makebox(0,0){\scalebox{3}{$*$}}}}
\put( -5 , 9 ){{\makebox(0,0){\scalebox{3}{$*$}}}}
\put( -4 , 9 ){{\makebox(0,0){\scalebox{3}{$*$}}}}
\put( -3 , 9 ){{\makebox(0,0){\scalebox{3}{$*$}}}}
\put( -2 , 9 ){{\makebox(0,0){\scalebox{3}{$*$}}}}
\put( -1 , 9 ){{\makebox(0,0){\scalebox{3}{$*$}}}}
\put( 0 , 9 ){{\makebox(0,0){\scalebox{3}{$*$}}}}
\put( 1 , 9 ){{\makebox(0,0){\scalebox{3}{$*$}}}}
\put( 2 , 9 ){{\makebox(0,0){\scalebox{3}{$*$}}}}
\put( 3 , 9 ){{\makebox(0,0){\scalebox{3}{$*$}}}}
\put( 4 , 9 ){{\makebox(0,0){\scalebox{3}{$*$}}}}
\put( 5 , 9 ){{\makebox(0,0){\scalebox{3}{$*$}}}}
\put( 6 , 9 ){{\makebox(0,0){\scalebox{3}{$*$}}}}
\put( 7 , 9 ){{\makebox(0,0){\scalebox{3}{$*$}}}}
\put( 8 , 9 ){{\makebox(0,0){\scalebox{3}{$*$}}}}
\put( 9 , 9 ){{\makebox(0,0){\scalebox{3}{$*$}}}}
\put( 10 , 9 ){{\makebox(0,0){\scalebox{3}{$*$}}}}
\put( 12 , 9 ){{\makebox(0,0){\scalebox{3}{$*$}}}}
\put( -10 , 10 ){{\makebox(0,0){\scalebox{3}{$*$}}}}
\put( -9 , 10 ){{\makebox(0,0){\scalebox{3}{$*$}}}}
\put( -8 , 10 ){{\makebox(0,0){\scalebox{3}{$*$}}}}
\put( -7 , 10 ){{\makebox(0,0){\scalebox{3}{$*$}}}}
\put( -6 , 10 ){{\makebox(0,0){\scalebox{3}{$*$}}}}
\put( -5 , 10 ){{\makebox(0,0){\scalebox{3}{$*$}}}}
\put( -4 , 10 ){{\makebox(0,0){\scalebox{3}{$*$}}}}
\put( -3 , 10 ){{\makebox(0,0){\scalebox{3}{$*$}}}}
\put( -2 , 10 ){{\makebox(0,0){\scalebox{3}{$*$}}}}
\put( -1 , 10 ){{\makebox(0,0){\scalebox{3}{$*$}}}}
\put( 0 , 10 ){{\makebox(0,0){\scalebox{3}{$*$}}}}
\put( 1 , 10 ){{\makebox(0,0){\scalebox{3}{$*$}}}}
\put( 2 , 10 ){{\makebox(0,0){\scalebox{3}{$*$}}}}
\put( 3 , 10 ){{\makebox(0,0){\scalebox{3}{$*$}}}}
\put( 4 , 10 ){{\makebox(0,0){\scalebox{3}{$*$}}}}
\put( 5 , 10 ){{\makebox(0,0){\scalebox{3}{$*$}}}}
\put( 6 , 10 ){{\makebox(0,0){\scalebox{3}{$*$}}}}
\put( 7 , 10 ){{\makebox(0,0){\scalebox{3}{$*$}}}}
\put( 8 , 10 ){{\makebox(0,0){\scalebox{3}{$*$}}}}
\put( 9 , 10 ){{\makebox(0,0){\scalebox{3}{$*$}}}}
\put( 10 , 10 ){{\makebox(0,0){\scalebox{3}{$*$}}}}
\put( -10 , 10 ){\circle*{4.5}}
\put( -9 , 10 ){\circle*{4.5}}
\put( -8 , 10 ){\circle*{4.5}}
\put( -7 , 10 ){\circle*{4.5}}
\put( -6 , 10 ){\circle*{4.5}}
\put( -5 , 10 ){\circle*{4.5}}
\put( -4 , 10 ){\circle*{4.5}}
\put( -3 , 10 ){\circle*{4.5}}
\put( -2 , 10 ){\circle*{4.5}}
\put( -1 , 10 ){\circle*{4.5}}
\put( 0 , 10 ){\circle*{4.5}}
\put( 1 , 10 ){\circle*{4.5}}
\put( 2 , 10 ){\circle*{4.5}}
\put( 3 , 10 ){\circle*{4.5}}
\put( 4 , 10 ){\circle*{4.5}}
\put( 5 , 10 ){\circle*{4.5}}
\put( 6 , 10 ){\circle*{4.5}}
\put( 7 , 10 ){\circle*{4.5}}
\put( 8 , 10 ){\circle*{4.5}}
\put( 9 , 10 ){\circle*{4.5}}
\put( 11 , 10 ){\scalebox{4}{${\bf x}_{( 10 , 3 )}= (1,\alpha^{5},\alpha,\alpha^{5},\alpha,1,1,\alpha,\alpha^{5},\alpha,1,\alpha^{5},\alpha^{5},1,\alpha,\alpha,\alpha^{5},1,\alpha,1,\alpha^{5}) $}}
\put( -12 , 11 ){{\makebox(0,0){\scalebox{3}{$*$}}}}
\put( -10 , 11 ){{\makebox(0,0){\scalebox{3}{$*$}}}}
\put( -9 , 11 ){{\makebox(0,0){\scalebox{3}{$*$}}}}
\put( -8 , 11 ){{\makebox(0,0){\scalebox{3}{$*$}}}}
\put( -7 , 11 ){{\makebox(0,0){\scalebox{3}{$*$}}}}
\put( -6 , 11 ){{\makebox(0,0){\scalebox{3}{$*$}}}}
\put( -5 , 11 ){{\makebox(0,0){\scalebox{3}{$*$}}}}
\put( -4 , 11 ){{\makebox(0,0){\scalebox{3}{$*$}}}}
\put( -3 , 11 ){{\makebox(0,0){\scalebox{3}{$*$}}}}
\put( -2 , 11 ){{\makebox(0,0){\scalebox{3}{$*$}}}}
\put( -1 , 11 ){{\makebox(0,0){\scalebox{3}{$*$}}}}
\put( 0 , 11 ){{\makebox(0,0){\scalebox{3}{$*$}}}}
\put( 1 , 11 ){{\makebox(0,0){\scalebox{3}{$*$}}}}
\put( 2 , 11 ){{\makebox(0,0){\scalebox{3}{$*$}}}}
\put( 3 , 11 ){{\makebox(0,0){\scalebox{3}{$*$}}}}
\put( 4 , 11 ){{\makebox(0,0){\scalebox{3}{$*$}}}}
\put( 5 , 11 ){{\makebox(0,0){\scalebox{3}{$*$}}}}
\put( 6 , 11 ){{\makebox(0,0){\scalebox{3}{$*$}}}}
\put( 7 , 11 ){{\makebox(0,0){\scalebox{3}{$*$}}}}
\put( 8 , 11 ){{\makebox(0,0){\scalebox{3}{$*$}}}}
\put( 10 , 11 ){{\makebox(0,0){\scalebox{3}{$*$}}}}
\put( -15 , 12 ){{\makebox(0,0){\scalebox{3}{$*$}}}}
\put( -12 , 12 ){{\makebox(0,0){\scalebox{3}{$*$}}}}
\put( -10 , 12 ){{\makebox(0,0){\scalebox{3}{$*$}}}}
\put( -9 , 12 ){{\makebox(0,0){\scalebox{3}{$*$}}}}
\put( -8 , 12 ){{\makebox(0,0){\scalebox{3}{$*$}}}}
\put( -7 , 12 ){{\makebox(0,0){\scalebox{3}{$*$}}}}
\put( -6 , 12 ){{\makebox(0,0){\scalebox{3}{$*$}}}}
\put( -5 , 12 ){{\makebox(0,0){\scalebox{3}{$*$}}}}
\put( -4 , 12 ){{\makebox(0,0){\scalebox{3}{$*$}}}}
\put( -3 , 12 ){{\makebox(0,0){\scalebox{3}{$*$}}}}
\put( -2 , 12 ){{\makebox(0,0){\scalebox{3}{$*$}}}}
\put( -1 , 12 ){{\makebox(0,0){\scalebox{3}{$*$}}}}
\put( 0 , 12 ){{\makebox(0,0){\scalebox{3}{$*$}}}}
\put( 1 , 12 ){{\makebox(0,0){\scalebox{3}{$*$}}}}
\put( 2 , 12 ){{\makebox(0,0){\scalebox{3}{$*$}}}}
\put( 3 , 12 ){{\makebox(0,0){\scalebox{3}{$*$}}}}
\put( 4 , 12 ){{\makebox(0,0){\scalebox{3}{$*$}}}}
\put( 5 , 12 ){{\makebox(0,0){\scalebox{3}{$*$}}}}
\put( 7 , 12 ){{\makebox(0,0){\scalebox{3}{$*$}}}}
\put( 8 , 12 ){{\makebox(0,0){\scalebox{3}{$*$}}}}
\put( 10 , 12 ){{\makebox(0,0){\scalebox{3}{$*$}}}}
\put( -15 , 13 ){{\makebox(0,0){\scalebox{3}{$*$}}}}
\put( -12 , 13 ){{\makebox(0,0){\scalebox{3}{$*$}}}}
\put( -11 , 13 ){{\makebox(0,0){\scalebox{3}{$*$}}}}
\put( -10 , 13 ){{\makebox(0,0){\scalebox{3}{$*$}}}}
\put( -9 , 13 ){{\makebox(0,0){\scalebox{3}{$*$}}}}
\put( -8 , 13 ){{\makebox(0,0){\scalebox{3}{$*$}}}}
\put( -7 , 13 ){{\makebox(0,0){\scalebox{3}{$*$}}}}
\put( -6 , 13 ){{\makebox(0,0){\scalebox{3}{$*$}}}}
\put( -5 , 13 ){{\makebox(0,0){\scalebox{3}{$*$}}}}
\put( -4 , 13 ){{\makebox(0,0){\scalebox{3}{$*$}}}}
\put( -3 , 13 ){{\makebox(0,0){\scalebox{3}{$*$}}}}
\put( -2 , 13 ){{\makebox(0,0){\scalebox{3}{$*$}}}}
\put( -1 , 13 ){{\makebox(0,0){\scalebox{3}{$*$}}}}
\put( 0 , 13 ){{\makebox(0,0){\scalebox{3}{$*$}}}}
\put( 1 , 13 ){{\makebox(0,0){\scalebox{3}{$*$}}}}
\put( 2 , 13 ){{\makebox(0,0){\scalebox{3}{$*$}}}}
\put( 3 , 13 ){{\makebox(0,0){\scalebox{3}{$*$}}}}
\put( 4 , 13 ){{\makebox(0,0){\scalebox{3}{$*$}}}}
\put( 5 , 13 ){{\makebox(0,0){\scalebox{3}{$*$}}}}
\put( 7 , 13 ){{\makebox(0,0){\scalebox{3}{$*$}}}}
\put( 8 , 13 ){{\makebox(0,0){\scalebox{3}{$*$}}}}
\put( -15 , 14 ){{\makebox(0,0){\scalebox{3}{$*$}}}}
\put( -14 , 14 ){{\makebox(0,0){\scalebox{3}{$*$}}}}
\put( -12 , 14 ){{\makebox(0,0){\scalebox{3}{$*$}}}}
\put( -11 , 14 ){{\makebox(0,0){\scalebox{3}{$*$}}}}
\put( -10 , 14 ){{\makebox(0,0){\scalebox{3}{$*$}}}}
\put( -9 , 14 ){{\makebox(0,0){\scalebox{3}{$*$}}}}
\put( -8 , 14 ){{\makebox(0,0){\scalebox{3}{$*$}}}}
\put( -7 , 14 ){{\makebox(0,0){\scalebox{3}{$*$}}}}
\put( -6 , 14 ){{\makebox(0,0){\scalebox{3}{$*$}}}}
\put( -5 , 14 ){{\makebox(0,0){\scalebox{3}{$*$}}}}
\put( -4 , 14 ){{\makebox(0,0){\scalebox{3}{$*$}}}}
\put( -3 , 14 ){{\makebox(0,0){\scalebox{3}{$*$}}}}
\put( -2 , 14 ){{\makebox(0,0){\scalebox{3}{$*$}}}}
\put( -1 , 14 ){{\makebox(0,0){\scalebox{3}{$*$}}}}
\put( 0 , 14 ){{\makebox(0,0){\scalebox{3}{$*$}}}}
\put( 1 , 14 ){{\makebox(0,0){\scalebox{3}{$*$}}}}
\put( 2 , 14 ){{\makebox(0,0){\scalebox{3}{$*$}}}}
\put( 3 , 14 ){{\makebox(0,0){\scalebox{3}{$*$}}}}
\put( 4 , 14 ){{\makebox(0,0){\scalebox{3}{$*$}}}}
\put( 5 , 14 ){{\makebox(0,0){\scalebox{3}{$*$}}}}
\put( 8 , 14 ){{\makebox(0,0){\scalebox{3}{$*$}}}}
\put( -17 , 15 ){{\makebox(0,0){\scalebox{3}{$*$}}}}
\put( -15 , 15 ){{\makebox(0,0){\scalebox{3}{$*$}}}}
\put( -14 , 15 ){{\makebox(0,0){\scalebox{3}{$*$}}}}
\put( -12 , 15 ){{\makebox(0,0){\scalebox{3}{$*$}}}}
\put( -11 , 15 ){{\makebox(0,0){\scalebox{3}{$*$}}}}
\put( -10 , 15 ){{\makebox(0,0){\scalebox{3}{$*$}}}}
\put( -9 , 15 ){{\makebox(0,0){\scalebox{3}{$*$}}}}
\put( -8 , 15 ){{\makebox(0,0){\scalebox{3}{$*$}}}}
\put( -7 , 15 ){{\makebox(0,0){\scalebox{3}{$*$}}}}
\put( -6 , 15 ){{\makebox(0,0){\scalebox{3}{$*$}}}}
\put( -5 , 15 ){{\makebox(0,0){\scalebox{3}{$*$}}}}
\put( -4 , 15 ){{\makebox(0,0){\scalebox{3}{$*$}}}}
\put( -3 , 15 ){{\makebox(0,0){\scalebox{3}{$*$}}}}
\put( -2 , 15 ){{\makebox(0,0){\scalebox{3}{$*$}}}}
\put( -1 , 15 ){{\makebox(0,0){\scalebox{3}{$*$}}}}
\put( 0 , 15 ){{\makebox(0,0){\scalebox{3}{$*$}}}}
\put( 1 , 15 ){{\makebox(0,0){\scalebox{3}{$*$}}}}
\put( 2 , 15 ){{\makebox(0,0){\scalebox{3}{$*$}}}}
\put( 3 , 15 ){{\makebox(0,0){\scalebox{3}{$*$}}}}
\put( 5 , 15 ){{\makebox(0,0){\scalebox{3}{$*$}}}}
\put( 8 , 15 ){{\makebox(0,0){\scalebox{3}{$*$}}}}
\put( -17 , 16 ){{\makebox(0,0){\scalebox{3}{$*$}}}}
\put( -15 , 16 ){{\makebox(0,0){\scalebox{3}{$*$}}}}
\put( -14 , 16 ){{\makebox(0,0){\scalebox{3}{$*$}}}}
\put( -13 , 16 ){{\makebox(0,0){\scalebox{3}{$*$}}}}
\put( -12 , 16 ){{\makebox(0,0){\scalebox{3}{$*$}}}}
\put( -11 , 16 ){{\makebox(0,0){\scalebox{3}{$*$}}}}
\put( -10 , 16 ){{\makebox(0,0){\scalebox{3}{$*$}}}}
\put( -9 , 16 ){{\makebox(0,0){\scalebox{3}{$*$}}}}
\put( -8 , 16 ){{\makebox(0,0){\scalebox{3}{$*$}}}}
\put( -7 , 16 ){{\makebox(0,0){\scalebox{3}{$*$}}}}
\put( -6 , 16 ){{\makebox(0,0){\scalebox{3}{$*$}}}}
\put( -5 , 16 ){{\makebox(0,0){\scalebox{3}{$*$}}}}
\put( -4 , 16 ){{\makebox(0,0){\scalebox{3}{$*$}}}}
\put( -3 , 16 ){{\makebox(0,0){\scalebox{3}{$*$}}}}
\put( -2 , 16 ){{\makebox(0,0){\scalebox{3}{$*$}}}}
\put( -1 , 16 ){{\makebox(0,0){\scalebox{3}{$*$}}}}
\put( 0 , 16 ){{\makebox(0,0){\scalebox{3}{$*$}}}}
\put( 1 , 16 ){{\makebox(0,0){\scalebox{3}{$*$}}}}
\put( 2 , 16 ){{\makebox(0,0){\scalebox{3}{$*$}}}}
\put( 3 , 16 ){{\makebox(0,0){\scalebox{3}{$*$}}}}
\put( 5 , 16 ){{\makebox(0,0){\scalebox{3}{$*$}}}}
\put( -17 , 17 ){{\makebox(0,0){\scalebox{3}{$*$}}}}
\put( -16 , 17 ){{\makebox(0,0){\scalebox{3}{$*$}}}}
\put( -15 , 17 ){{\makebox(0,0){\scalebox{3}{$*$}}}}
\put( -14 , 17 ){{\makebox(0,0){\scalebox{3}{$*$}}}}
\put( -13 , 17 ){{\makebox(0,0){\scalebox{3}{$*$}}}}
\put( -12 , 17 ){{\makebox(0,0){\scalebox{3}{$*$}}}}
\put( -11 , 17 ){{\makebox(0,0){\scalebox{3}{$*$}}}}
\put( -10 , 17 ){{\makebox(0,0){\scalebox{3}{$*$}}}}
\put( -9 , 17 ){{\makebox(0,0){\scalebox{3}{$*$}}}}
\put( -8 , 17 ){{\makebox(0,0){\scalebox{3}{$*$}}}}
\put( -7 , 17 ){{\makebox(0,0){\scalebox{3}{$*$}}}}
\put( -6 , 17 ){{\makebox(0,0){\scalebox{3}{$*$}}}}
\put( -5 , 17 ){{\makebox(0,0){\scalebox{3}{$*$}}}}
\put( -4 , 17 ){{\makebox(0,0){\scalebox{3}{$*$}}}}
\put( -3 , 17 ){{\makebox(0,0){\scalebox{3}{$*$}}}}
\put( -2 , 17 ){{\makebox(0,0){\scalebox{3}{$*$}}}}
\put( -1 , 17 ){{\makebox(0,0){\scalebox{3}{$*$}}}}
\put( 0 , 17 ){{\makebox(0,0){\scalebox{3}{$*$}}}}
\put( 1 , 17 ){{\makebox(0,0){\scalebox{3}{$*$}}}}
\put( 2 , 17 ){{\makebox(0,0){\scalebox{3}{$*$}}}}
\put( 3 , 17 ){{\makebox(0,0){\scalebox{3}{$*$}}}}
\put( -17 , 17 ){\circle*{4.5}}
\put( -16 , 17 ){\circle*{4.5}}
\put( -15 , 17 ){\circle*{4.5}}
\put( -14 , 17 ){\circle*{4.5}}
\put( -13 , 17 ){\circle*{4.5}}
\put( -12 , 17 ){\circle*{4.5}}
\put( -11 , 17 ){\circle*{4.5}}
\put( -10 , 17 ){\circle*{4.5}}
\put( -9 , 17 ){\circle*{4.5}}
\put( -8 , 17 ){\circle*{4.5}}
\put( -7 , 17 ){\circle*{4.5}}
\put( -6 , 17 ){\circle*{4.5}}
\put( -5 , 17 ){\circle*{4.5}}
\put( -4 , 17 ){\circle*{4.5}}
\put( -3 , 17 ){\circle*{4.5}}
\put( -2 , 17 ){\circle*{4.5}}
\put( -1 , 17 ){\circle*{4.5}}
\put( 0 , 17 ){\circle*{4.5}}
\put( 1 , 17 ){\circle*{4.5}}
\put( 2 , 17 ){\circle*{4.5}}
\put( 4 , 17 ){\scalebox{4}{${\bf x}_{( 17 , 3 )}= (1,\alpha^{6},\alpha^{4},\alpha^{6},\alpha^{4},1,1,\alpha^{4},\alpha^{6},\alpha^{4},1,\alpha^{6},\alpha^{6},1,\alpha^{4},\alpha^{4},\alpha^{6},1,\alpha^{4},1,\alpha^{6}) $}}
\put( -19 , 18 ){{\makebox(0,0){\scalebox{3}{$*$}}}}
\put( -17 , 18 ){{\makebox(0,0){\scalebox{3}{$*$}}}}
\put( -16 , 18 ){{\makebox(0,0){\scalebox{3}{$*$}}}}
\put( -15 , 18 ){{\makebox(0,0){\scalebox{3}{$*$}}}}
\put( -14 , 18 ){{\makebox(0,0){\scalebox{3}{$*$}}}}
\put( -13 , 18 ){{\makebox(0,0){\scalebox{3}{$*$}}}}
\put( -12 , 18 ){{\makebox(0,0){\scalebox{3}{$*$}}}}
\put( -11 , 18 ){{\makebox(0,0){\scalebox{3}{$*$}}}}
\put( -10 , 18 ){{\makebox(0,0){\scalebox{3}{$*$}}}}
\put( -9 , 18 ){{\makebox(0,0){\scalebox{3}{$*$}}}}
\put( -8 , 18 ){{\makebox(0,0){\scalebox{3}{$*$}}}}
\put( -7 , 18 ){{\makebox(0,0){\scalebox{3}{$*$}}}}
\put( -6 , 18 ){{\makebox(0,0){\scalebox{3}{$*$}}}}
\put( -5 , 18 ){{\makebox(0,0){\scalebox{3}{$*$}}}}
\put( -4 , 18 ){{\makebox(0,0){\scalebox{3}{$*$}}}}
\put( -3 , 18 ){{\makebox(0,0){\scalebox{3}{$*$}}}}
\put( -2 , 18 ){{\makebox(0,0){\scalebox{3}{$*$}}}}
\put( -1 , 18 ){{\makebox(0,0){\scalebox{3}{$*$}}}}
\put( 0 , 18 ){{\makebox(0,0){\scalebox{3}{$*$}}}}
\put( 1 , 18 ){{\makebox(0,0){\scalebox{3}{$*$}}}}
\put( 3 , 18 ){{\makebox(0,0){\scalebox{3}{$*$}}}}
\put( -22 , 19 ){{\makebox(0,0){\scalebox{3}{$*$}}}}
\put( -19 , 19 ){{\makebox(0,0){\scalebox{3}{$*$}}}}
\put( -17 , 19 ){{\makebox(0,0){\scalebox{3}{$*$}}}}
\put( -16 , 19 ){{\makebox(0,0){\scalebox{3}{$*$}}}}
\put( -15 , 19 ){{\makebox(0,0){\scalebox{3}{$*$}}}}
\put( -14 , 19 ){{\makebox(0,0){\scalebox{3}{$*$}}}}
\put( -13 , 19 ){{\makebox(0,0){\scalebox{3}{$*$}}}}
\put( -12 , 19 ){{\makebox(0,0){\scalebox{3}{$*$}}}}
\put( -11 , 19 ){{\makebox(0,0){\scalebox{3}{$*$}}}}
\put( -10 , 19 ){{\makebox(0,0){\scalebox{3}{$*$}}}}
\put( -9 , 19 ){{\makebox(0,0){\scalebox{3}{$*$}}}}
\put( -8 , 19 ){{\makebox(0,0){\scalebox{3}{$*$}}}}
\put( -7 , 19 ){{\makebox(0,0){\scalebox{3}{$*$}}}}
\put( -6 , 19 ){{\makebox(0,0){\scalebox{3}{$*$}}}}
\put( -5 , 19 ){{\makebox(0,0){\scalebox{3}{$*$}}}}
\put( -4 , 19 ){{\makebox(0,0){\scalebox{3}{$*$}}}}
\put( -3 , 19 ){{\makebox(0,0){\scalebox{3}{$*$}}}}
\put( -2 , 19 ){{\makebox(0,0){\scalebox{3}{$*$}}}}
\put( 0 , 19 ){{\makebox(0,0){\scalebox{3}{$*$}}}}
\put( 1 , 19 ){{\makebox(0,0){\scalebox{3}{$*$}}}}
\put( 3 , 19 ){{\makebox(0,0){\scalebox{3}{$*$}}}}
\put( -22 , 20 ){{\makebox(0,0){\scalebox{3}{$*$}}}}
\put( -19 , 20 ){{\makebox(0,0){\scalebox{3}{$*$}}}}
\put( -18 , 20 ){{\makebox(0,0){\scalebox{3}{$*$}}}}
\put( -17 , 20 ){{\makebox(0,0){\scalebox{3}{$*$}}}}
\put( -16 , 20 ){{\makebox(0,0){\scalebox{3}{$*$}}}}
\put( -15 , 20 ){{\makebox(0,0){\scalebox{3}{$*$}}}}
\put( -14 , 20 ){{\makebox(0,0){\scalebox{3}{$*$}}}}
\put( -13 , 20 ){{\makebox(0,0){\scalebox{3}{$*$}}}}
\put( -12 , 20 ){{\makebox(0,0){\scalebox{3}{$*$}}}}
\put( -11 , 20 ){{\makebox(0,0){\scalebox{3}{$*$}}}}
\put( -10 , 20 ){{\makebox(0,0){\scalebox{3}{$*$}}}}
\put( -9 , 20 ){{\makebox(0,0){\scalebox{3}{$*$}}}}
\put( -8 , 20 ){{\makebox(0,0){\scalebox{3}{$*$}}}}
\put( -7 , 20 ){{\makebox(0,0){\scalebox{3}{$*$}}}}
\put( -6 , 20 ){{\makebox(0,0){\scalebox{3}{$*$}}}}
\put( -5 , 20 ){{\makebox(0,0){\scalebox{3}{$*$}}}}
\put( -4 , 20 ){{\makebox(0,0){\scalebox{3}{$*$}}}}
\put( -3 , 20 ){{\makebox(0,0){\scalebox{3}{$*$}}}}
\put( -2 , 20 ){{\makebox(0,0){\scalebox{3}{$*$}}}}
\put( 0 , 20 ){{\makebox(0,0){\scalebox{3}{$*$}}}}
\put( 1 , 20 ){{\makebox(0,0){\scalebox{3}{$*$}}}}
\end{picture}}
\\
\phantom{M}\begin{minipage}{.95\textwidth}{\footnotesize 
\begin{picture}(.25,.25)
\put(0.1,0.1){\makebox(0,0){\scalebox{ 0.800000000000000 }{$*$}}}\end{picture}
at $(a,\beta_1)$ if $a\in H_{(\beta_1Q_1+\beta_2Q_2)}^*$\\
\begin{picture}(.25,.25)
\put(0.1,0.1){\circle*{.1}}
\end{picture}
at $(a,\beta_1)$ if $C_{{\mathcal L}}(D,aP_\infty+\beta_1Q_1+\beta_2Q_2)$
is in an isometry-dual flag 
}\end{minipage}\end{minipage}
\end{center}\end{figure}
}

\title{Self-orthogonal flags of codes and translation of flags of algebraic geometry codes}

\author{Maria Bras-Amorós, Alonso S. Castellanos, Luciane Quoos}
\date{\today}

\begin{document}
\maketitle

\section*{Abstract}

A flag $C_0 \subsetneq C_1 \cdots \subsetneq C_s \subsetneq \fq^n $ of linear codes is said to be self-orthogonal if the duals of the codes in the flag satisfy $C_{i}^\perp=C_{s-i}$, and it is said to satisfy the isometry-dual property with respect to an isometry vector ${\bf x}$ if $C_i^\perp={\bf x} C_{s-i}$ for $i=1, \dots, s$.
We characterize complete (i.e. $s=n$) flags with the isometry-dual property by means of the existence of a word with non-zero coordinates in a certain linear subspace of $\fq^n$. For flags of algebraic geometry (AG) codes we prove a so-called translation property of isometry-dual flags and give a construction of complete self-orthogonal flags, providing examples of self-orthogonal flags over some maximal function fields.  At the end we characterize the divisors giving the isometry-dual property and the related isometry vectors showing that for each function field there is only a finite number of isometry vectors and that they are related by cyclic repetitions.

\section{Introduction}
Let $\fq$ be the finite field with $q$ elements. A {\em linear code} $C$ of {\em length} $n$ and {\em dimension} $k$ is a $\fq$-linear subspace of $\fq^n, n\geq 1$, where $k$ is also the dimension of $C$ as a $\fq$-linear subspace. A {\it flag of codes} is a sequence $C_0\subsetneq C_1\dots\subsetneq C_s$ of codes in $\fq^n$. In the last years flags of codes have been investigated due their applications to network coding, see \cite{LNV2018}. Also a notion of distance of the flag variety was introduced to study error correcting codes. This was investigated in \cite{CY2023, ANAX2023} and \cite{K2021}. Recent research on isometry-dual flags of algebraic geometry codes can be found in \cite{BDH2020}. In this work we continue our previous investigation on complete flags given by algebraic geometry codes in \cite{BCQ2022} and \cite{BCQ2023}.

The {\em dual} code $C^\perp$ of $C$ is the orthogonal complement of $C$ in $\fq^n$ with the standard inner product of $\fq^n$. A code is said to be {\em self-dual} if $C=C^\perp$.
The restrictive condition of self-dual codes can be relaxed by investigating orthogonality in flags of linear codes $C_0 \subsetneq C_1 \subsetneq \cdots \subsetneq C_s$.
Let us use $*$ for the componentwise product of vectors and  $\cdot$ for the inner product in $\fq^n $.
A flag $C_0\subsetneq C_1\dots\subsetneq C_s$ 
is said to satisfy the {\it isometry-dual property} (first introduced in \cite{GMRT2011}) if there exists a vector ${\bf x}\in (\mathbb{F}_q^*)^n$ such that $C_i$ is {\bf x}-isometric to $C_{s-i}^\bot$ for all $i=0,\ldots, s$, that is, $C_{s-i}^\perp={\bf x} * C_{i}$. In this case, ${\bf x}$ is called an {\em isometry vector}.
We will sometimes specify the isometry vector by saying that $C_0\subsetneq C_1\dots\subsetneq C_s$ satisfies the ${\bf x}$-isometry-dual property.
The flag of codes is said to be {\it self-orthogonal} if it satisfies the isometry-dual property with respect to the constant vector ${\bf 1}$. 
Notice that, if $(C_i)_{i=0, \dots , n}$ is self-orthogonal, then we have $C_{n-i}^\perp=C_i \subsetneq C_{n-i}$ for $i=1,\dots, \lfloor{\frac{n}2}\rfloor$. That is, the code $C_{n-i}$ is self-orthogonal.

Function fields with many rational places are appreciated in coding theory since they allow the construction of codes with good parameters.
A function field over $\fqs $ is called a {\it maximal function field} if its number of rational places attains the Hasse-Weil upper bound $q+1+2gq$. Constructions of AG codes over maximal function fields have been a proficuous object of research in the last decades, see \cite{XC2002,MST2008,B2006, AM2020} and references in there.

We now define the flags of algebraic geometry codes we consider in this work. This follows the investigation in \cite{BCQ2023}.

Let $\cF$  be a function field of genus $g$ over $\mathbb{F}_q$ and let $P,P_1,\ldots,P_n, Q_1,\ldots,Q_t$ be pairwise distinct rational places of $\cF$.
Given a tuple $\b=(\b_1, \dots, \b_t)$ of integers we denote by ${\bf G_{\b}}$ the divisor $\sum_{i=1}^t\b_iQ_i$ and $D=P_1+ \cdots +P_n$.  
For $a\in{\mathbb Z}$, define the (linear) algebraic geometry code 
\begin{equation}\label{code}
C_\mathcal L(D,aP+{\bf G}_\b) =\{(f(P_1),\dots,f(P_n)) \mid f\in{\mathcal L}(aP+{\bf G}_\b)\},
\end{equation}
where ${\mathcal L}(G)=\{z\in{\mathcal F} \mid (z)\geq -G\}$ denotes the Riemann-Roch space associated to a divisor $G$. 

For $s$ an integer with $0 \leq s\leq n$ consider
$a_1,\dots,a_s \in \Z$  such that $a_0 <a_1< \cdots < a_s$, and $\sum_{i=1}^t \b_i +a_0 \geq 0$. 
Define the flag $S_\b^{a_0,\dots,a_s}$ by
\begin{equation}\label{flag}
S^{a_0,\dots,a_s}_\b:\, C_\mathcal L(D, a_0P+{\bf G}_\b)\subsetneq C_\mathcal L(D, a_1P+{\bf G}_\b))\subsetneq \dots \subsetneq C_\mathcal L(D, a_sP+{\bf G}_\b).
\end{equation}
Fixed a tuple $\b=(\b_1, \dots, \b_t)$ of integers consider the sets
\begin{align*}
&H_{\bf \b}=\{a\in\Z  \, \mid \,  \ell((a-1)P+{\bf G_{\b}})\neq \ell(aP+{\bf G_{\b}}) \}, \text{ and }\\
&\\
&H_{\bf \b}^*=\{a\in\Z \, \mid \, C_\mathcal L(D, (a-1)P+{\bf G_\b})\neq C_\mathcal L(D, aP+{\bf G_\b}) \}.
\end{align*}
We notice that, if $a$ in $H_{\bf \b}^*$, then $a+\sum_{i=1}^t\b_iQ_i\geq 0$. It holds that 
$a \in H_{\bf \b}^*$ if and only if
$$\dim C_\mathcal L(D, aP+{\bf G_\b})-\dim C_\mathcal L(D, (a-1)P+{\bf G_\b})=1.$$

The set $H_{\b}^*$ has $n$ elements.
See \cite{BCQ2023} for more results related to $H_\b$ and $H_\b^*$.
Let $a_0(\b)=\min H_{\b}^*-1$. We define
\begin{equation}\label{defS}S_\b=S_\b^{a_0(\b)\cup {H}_{\b}^*}.\end{equation}
Flags of codes that contain $n+1$ codes of length $n$, all of them of different dimension, are called {\em complete flags of codes}.
We notice that $S_\b$ is, indeed, a complete flag of codes.
We remark also that in the case of complete flags of codes, the isometry vector is unique up to multiplication by non-zero scalars. We say that a vector is {\em normalized} if its first nonzero component is $1$. We will always consider normalized isometry vectors when referring to {\em the} isometry vector.

In Section~\ref{s:bb} we use the notion of a building base of a complete flag of codes and characterize complete flags of codes satisfying the isometry dual property in terms of words of maximum weight in the null space of a given linear space. In particular, given a flag of codes, we give a set of parity checks of a new code such that the flag satisfies the isometry-dual property if and only if the new code has a word with no zero coordinates. The study of codewords with maximum weight has been object of research in \cite{BB2013} and \cite{O2016}.

In Section~\ref{s:translation} we see that, under the assumption of the existence of a function with a certain divisor, if a vector $\b \in \Z^t$ is such that $S_\b$ satisfies the isometry-dual property, then so is a translation of it, determined by the divisor of the function. Such functions always exist, for instance, for the broad class of curves derived from Kummer extensions. 

In Section~\ref{s:sufficientcondition} we prove a sufficient condition under which there exists a set of vectors $\b \in \Z^t$ such that $S_\b$ is self-orthogonal. In this case we give an explicit construction of the flag.

In Section~\ref{s:charact}, under the assumption of the existence of a vector $\b\in{\mathbb Z}^t$ such that $S_\b$ satisfies the isometry-dual property, 
we give a characterization of the whole set of vectors $\gamma\in{\mathbb Z}^t$ such that $S_\gamma$ satisfies the isometry-dual property. Furthermore, we describe the corresponding isometry vectors. In particular, we see that there is a cyclic repetition of the isometry vectors and give an upper bound on their number.

\section{The isometry-dual property in terms of building bases}
\label{s:bb}

Suppose $C$ is a $\fq$-linear code with length $n$, dimension $k$, and
linear basis given by 
$u_1,u_2,\dots ,u_k$ in $ \fq^n$.
That is, $C$ is the linear space spanned by $u_1,u_2,\dots ,u_k$.
Let $\tilde C$ be a second linear code with the same length $n$, dimension $r=n-k$, and with
linear basis
$v_1,v_2,\dots ,v_r \in \fq^n$.
Then $C={\tilde C}^\perp$ if and only if $u_i\cdot v_j=0$ for all $i, j$ with $1\leq i\leq k$ and $1\leq j\leq r$.

For a vector ${\bf x}$ in $ ({\fq^*})^n$, a generator matrix of the equivalent code  ${\bf x}* C$ has rows ${\bf x}* u_1, {\bf x}* u_2,\dots,{\bf x}* u_k$. 
Consequently, 
$  {\tilde C}^\perp=  {\bf x} * C$ if and only if for all $i, j$ with $1\leq i\leq k$ and $1\leq j\leq r$ we have
\begin{equation}\label{conddual}
({\bf x}* u_i)\cdot  v_j=0 
\Longleftrightarrow
\sum_{s=1}^n{\bf x}_s(u_i)_s(v_j)_s=0
\Longleftrightarrow
(u_i* v_j)\cdot {\bf x}=0.
\end{equation}

Now we want to apply these results for flags satisfying the ${\bf x}$-isometry-dual property. For $0\leq i\leq m$
let $C_i$ be a linear code with length $n$, dimension $k_i$, and
linear basis
$u^i_1,u^i_2,\dots u^i_{k_i}$ in $\fq^n$.
Then, the sequence of codes 
$$(C_i)_{i=0,\dots , m}: \quad C_0 \subsetneq C_2 \subsetneq \cdots \subsetneq C_m$$ 
satisfies the isometry dual property if and only if there exists  a vector ${\bf x} \in (\fq^*)^n$ such that 
$C_{m-i}^\perp= {\bf x} * C_i$ for  $i=0, \dots, m$.
That is, if and only if there exists ${\bf x} \in (\fq^*)^n$ such that 
$$(u^i_\ii* u^{m-i}_\jj)\cdot {\bf x}=0, \quad \text{ for all } 0\leq i\leq m,\quad 1\leq \ii\leq k_i, \quad 1\leq \jj\leq k_{m-i} $$
by \eqref{conddual}.
That is, the linear system on $x$ in $\fq^n$
$$(u^i_\ii* u^{m-i}_\jj)\cdot x=0$$
has a solution {\bf x} with nonzero coordinates.
We have proved the following lemma.
\begin{lemma}\label{lemma:charactx}
  For $0\leq i\leq m$
  let $C_i$ be a linear code with length $n$, dimension $k_i$, and
  linear basis
  $u^i_1,u^i_2,\dots u^i_{k_i}$ in $\fq^n$.
  A vector $x\in{\mathbb F}_q^n$ (not necessarily with nonzero coordinates)
  satisfies 
  $C_{m-i}^\perp=x* C_i$ for all $i$ with $0\leq i\leq m$ if and only if $x$ belongs to the nullspace of
the matrix with rows $$\left\{u^i_{\ii}* u^{m-i}_{\jj}\right\}_{\begin{array}{l}
0\leq i\leq m\\
1\leq \ii\leq k_i\\
1\leq \jj\leq k_{m-i}\end{array}}.$$
\end{lemma}
\begin{definition}
 We say that a set of vectors $\{v_1,\dots,v_n\}$ of ${\mathbb F}_q^n$ is a {\em building} basis of a complete flag of codes $\{0\}=C_0\subsetneq C_1\subsetneq \dots\subsetneq C_n=\fq^n$ if
  $v_{i}\in C_{i}\setminus C_{i-1}$ for every $i=1, \dots, n$.
  \end{definition}
In particular, a building basis of a complete flag of codes is a basis of ${\mathbb F}_q^n$, where $n$ is the code length and, for any
$i$ from $1$ to $n$, we have that $C_i$ is the vector space generated by $v_1,\dots, v_{i}$.

For complete flags of codes, Lemma~\ref{lemma:charactx} can be restated as follows, in terms of building bases.

\begin{theorem}\label{lemma:charactxrows}
  Let $\{0\}=C_0\subsetneq C_1\subsetneq \dots\subsetneq C_n=\fq^n$ be a complete flag with building basis $v_1,\dots, v_n$.
  Then, the flag of codes satisfies the isometry-dual property if and only if the code corresponding to the null space of 
$$\left\{v_{\ii}* v_{\jj}\right\}_{\begin{array}{l}
2\leq \ii+\jj\leq n+1
\end{array}}$$
has a codeword ${\bf x}$ with weight $n$. Each of the codewords ${\bf x}$ of this null space gives the ${\bf x}$-isometry-dual property.
\end{theorem}
\begin{proof}
Using the notation of Lemma~\ref{lemma:charactx} we are going to show that the flag of codes satisfies the isometry-dual property with respect to a vector ${\bf x} \in (\fq^*)^n$ if and only if the vector ${\bf x}$ belongs to the code defined as the null space of the matrix with rows 
$$\left\{v_{\ii}* v_{\jj}\right\}_{\begin{array}{l}
2\leq \ii+\jj\leq n
\end{array}}.$$
  First of all, notice that $C_i$ is the vector space spanned by $v_1,\dots, v_{i}$. So, we have that $k_i=i$, $k_{n-i}=n-i$, and $C_i$ is generated by
  $u^i_1=v_1,u^i_2=v_2,\dots, u^i_{i}=v_{i}$, while
  $C_{n-i}$ is generated by
  $u^{n-i}_1=v_1,u^{n-i}_2=v_2,\dots, u^{n-i}_{n-i}=v_{n-i}$. That is, 
  $u^i_\ii=v_\ii$ for any $\ii$ with $1\leq \ii\leq i$ and
  and $u^{n-i}_\jj=v_\jj$ for any $\jj$ with $1\leq\jj\leq n-i$.
Then, the lemma is a consequence of Lemma \ref{lemma:charactx} and the fact that
\begin{eqnarray*}
\left\{u^i_{\ii}* u^{n-i}_{\jj}\right\}_{\begin{array}{l}1\leq i\leq n\\
1\leq \ii\leq k_i\\
1\leq \jj\leq k_{n-i}\end{array}}&=&
\left\{v_{\ii}* v_{\jj}\right\}_{\begin{array}{l}
1\leq i\leq n \\
1\leq \ii\leq i\\
1\leq \jj\leq n-i\end{array}}\\&=&
\left\{v_{\ii}* v_{\jj}\right\}_{\begin{array}{l}
2\leq \ii+\jj\leq n
\end{array}}.
\end{eqnarray*}
  \end{proof}

\section{Translation of the isometry-dual property}  
\label{s:translation}

\begin{center}
\graphkleinD 
\end{center}

We start this section with the investigation of a flag over the Klein quartic defined over the finite field ${\mathbb F}_8={\mathbb F}_2(\alpha)$ with $\alpha^3+\alpha+1=0$  by the projective equation $X^3Y+Y^3Z+XZ^3=0$. This curve has genus $3$ and $24$ rational places. This curve was introduced by F. Klein \cite{K1879} and has been studied thoroughly. Fix the rational places $P=(1 : 0 : 0)$, $Q_1=(0 : 1 : 0)$ and $Q_2=(0 : 0 : 1)$.
Let $D$ be the divisor defined by the sum of the other rational places in the function field, that is, 
\begin{align*}
D&=P_1+\dots+P_{21}\\
&=(1 : \alpha : 1)+(\alpha^5 : \alpha^6 : 1)+(\alpha : \alpha^2 : 1)+(1 : \alpha^2 : 1)+(\alpha^3 : \alpha^5 : 1)
\\&+(\alpha^2 : \alpha^4 : 1)+(\alpha^4 : 1 : 1)+(\alpha^5 : \alpha : 1)+(\alpha^2 : \alpha^5 : 1)+(1 : \alpha^4 : 1)\\
&+(\alpha^6 : \alpha^3 : 1)+(\alpha^4 : \alpha : 1)+(\alpha^6 : \alpha^4 : 1)+(\alpha : \alpha^6 : 1)+(\alpha^2 : 1 : 1)\\
&+(\alpha^4 : \alpha^3 : 1)+(\alpha : 1 : 1)+(\alpha^3  : \alpha^2 : 1)+(\alpha^6 : \alpha^6 : 1)+(\alpha^5 : \alpha^5 : 1)\\
&+(\alpha^3 : \alpha^3 : 1).
\end{align*}
In this case $f= Y^3/X$ has divisor equal to $7P-7Q_1$ and the corresponding vector $\tau=(f(P_i)/f(P_1))_{P_i\in \supp D}$ is $$\tau= (1,\alpha^{3},\alpha^{2},\alpha^{3},\alpha^{2},1,1,\alpha^{2},\alpha^{3},\alpha^{2},1,\alpha^{3},\alpha^{3},1,\alpha^{2},\alpha^{2},\alpha^{3},1,\alpha^{2},1,\alpha^{3}) .$$

We observe in the graphic over the Klein curve, see Figure \ref{klein} that $S_{(-4,3)}$ satisfies the isometry dual property with respect to the normalized vector
$${\bf x}_{( -4 , 3 )}= (1,\alpha^{3},\alpha^{2},\alpha^{3},\alpha^{2},1,1,\alpha^{2},\alpha^{3},\alpha^{2},1,\alpha^{3},\alpha^{3},1,\alpha^{2},\alpha^{2},\alpha^{3},1,\alpha^{2},1,\alpha^{3}) .$$
On the other hand we observe that $S_{(3,3)}$ satisfies the isometry dual property with respect to the vector
$${\bf x}_{( 3 , 3 )}= (1,\alpha^{4},\alpha^{5},\alpha^{4},\alpha^{5},1,1,\alpha^{5},\alpha^{4},\alpha^{5},1,\alpha^{4},\alpha^{4},1,\alpha^{5},\alpha^{5},\alpha^{4},1,\alpha^{5},1,\alpha^{4}).$$

It holds that 
$${\bf x}_{ (3 , 3) }=\left(\frac{({\bf x}_{(-4 , 3)})_i}{\tau_i^2}\right)_{i\in\{1,\dots,n\}}.$$

Similarly, 
${\bf x}_{ (10 , 3) }=\left(\frac{({\bf x}_{(3 , 3)})_i}{\tau_i^2}\right)_{i\in\{1,\dots,n\}}$
and
${\bf x}_{ (17 , 3) }=\left(\frac{({\bf x}_{(10 , 3)})_i}{\tau_i^2}\right)_{i\in\{1,\dots,n\}}.$

This translation behavior of the vectors $(-4,3)$, $(3,3)$, $(10,3)$,\dots by the vector $(7,0)$, as well as the relationship between the corresponding isometry-vectors ${\bf x}_{(-4,3)}$, ${\bf x}_{(3,3)}$, ${\bf x}_{(10,3)}$,\dots with respect to the vector $\tau$ is proved in next theorem.

\begin{theorem}\label{t:translationplus}
  Let $\cF/\fq$ be a function field and $P, P_1, \dots, P_n, Q_1, \dots Q_t$ pairwise distinct rational places.
Fix a $t$-tuple $(\b_1, \dots, \b_t)$ in $\Z^t$, and consider $S_\beta$ as defined in \eqref{defS}.
Suppose there exists a function $f$ in $ \cF$ with divisor 
\begin{equation}\label{functionf}
(f)=uP-u_1Q_1-\dots-u_tQ_t
\end{equation} 
for some $u \geq 1$ and $u_i \geq 0$ for $i=1, \dots, t$ and consider  the vector $${\bf \tau}=(1,f(P_2)/f(P_1),\dots, f(P_n)/f(P_1)).$$
Then
\begin{enumerate}[i)]
\item
  $H_{(\beta_1+u_1,\dots,\beta_t+u_t)}=\{a-u\mid a\in H_{\b}\}$,
  \item
  $H_{(\beta_1+u_1,\dots,\beta_t+u_t)}^*=\{a-u\mid a\in H_{\b}^*\}$,
\item 
The flag $S_{(\beta_1,\dots,\beta_t)}$ satisfies the isometry-dual property if and only if the flag $S_{(\beta_1+u_1,\dots,\beta_t+u_t)}$ satisfies the isometry-dual property.
In this case, we also have the following relation between the corresponding isometry vectors 
${\bf x}_{(\beta_1+u_1,\dots,\beta_t+u_t)}$ and ${\bf x}_{(\beta_1,\dots,\beta_t)}$.
\begin{equation}
({\bf x}_{(\beta_1+u_1,\dots,\beta_t+u_t)})_i=\frac{({\bf x}_{(\beta_1,\dots,\beta_t)})_i}{\tau_i^2} \text{ for }  i\in\{1,\dots, n\}.
\end{equation}
\end{enumerate}
\end{theorem}
\begin{proof}
First of all, since $(f)=uP-u_1Q_1-\dots-u_tQ_t$, notice that $\tau$ has no zero components. 
  \begin{enumerate}[i)]
  \item
The set $\{w_1,\dots, w_\ell\}$ is a basis of $\cL(aP+{\bf G}_\b)$ if and only if $\{f w_1,\dots,f w_\ell\}$ is a basis of $\cL((a-u)P+{\bf G}_{(\beta_1+u_1,\dots,\beta_t+u_t)})$. Hence,
$\ell(aP+{\bf G}_\b)=\ell((a-u)P+{\bf G}_{(\beta_1+u_1,\dots,\beta_t+u_t)})$ and 
      $H_{(\beta_1+u_1,\dots,\beta_t+u_t)}=\{a-u\mid a\in H_{(\beta_1,\dots,\beta_t)}\}$.  
\item
  We start by noticing that for $\phi \in {\mathcal F}$, the vector $((f \phi)(P_1),\dots,(f \phi)(P_n))$ is a constant multiple of the vector $(\tau_1\phi(P_1),\dots,\tau_n \phi (P_n))$. 
      
From the first item we have that if
$a\in H_{(\beta_1,\dots,\beta_t)}$
      then $a-u\in H_{(\beta_1+u_1,\dots,\beta_t+u_t)}$.
      Let $\ell=\ell(aP+G_\b)=\ell((a-u)P+{\bf G}_{(\beta_1+u_1,\dots,\beta_t+u_t)})$ and 
      $w_1,\dots, w_\ell$  be a basis of $\cL(aP+{\bf G}_\b)$ such that $w_1,\dots, w_{\ell-1}$ is a basis of $\cL((a-1)P+{\bf G}_\b)$.
      
      Now,
      $C_{\mathcal L}(D,aP+{\bf G}_\b)=C_{\mathcal L}(D,(a-1)P+{\bf G}_\b)$ if and only if there exist $\lambda_1,\dots,\lambda_{\ell-1}\in{\mathbb F}_q$ such that
      $$(w_\ell(P_1),\dots,w_\ell(P_n))=\sum_{i=1}^{\ell -1} \lambda_i(w_i(P_1),\dots,w_i(P_n)),$$
      which is equivalent to 
      $$(\tau_1w_\ell(P_1),\dots,\tau_nw_\ell(P_n))= \sum_{i=1}^{\ell-1}\lambda_i(\tau_1w_i(P_1),\dots,\tau_nw_i(P_n)),$$
      which in turn is equivalent to 
      $$C_{\mathcal L}(D,(a-u)P+{\bf G}_{(\beta_1+u_1,\dots,\beta_t+u_t)})= C_{\mathcal L}(D,(a-u-1)P+{\bf G}_{(\beta_1+u_1,\dots,\beta_t+u_t)}),$$ 
      since $w_1, \dots, w_\ell$ is a basis of $\cL(aP+{\bf G}_{\bf \b})$ if and only if $f w_1, \dots, f w_\ell$ is a basis of $\cL((a-u)P+{\bf G}_{(\beta_1+u_1,\dots,\beta_t+u_t)})$.

      So,
      $H^*_{(\beta_1+u_1,\dots,\beta_t+u_t)}=\{a-u\mid a\in H^*_{\b}\}$.        
    \item Suppose now that $\varphi_1,\dots,\varphi_n$ are functions in $\cF$ such that
      any code of the form
      $C_{\mathcal L}(D,aP+{\bf G}_{\bf \b})$
      is generated by the vectors
      $$(\varphi_1(P_1),\dots,\varphi_1(P_n)),\dots,(\varphi_{k}(P_1),\dots,\varphi_{k}(P_n))$$
      for some $k$.
That is, $\{(\varphi_1(P_1),\dots,\varphi_1(P_n)),\dots,(\varphi_{n}(P_1),\dots,\varphi_{n}(P_n))\}$ is a building basis of $S_\b$.
Then, by Lemma~\ref{lemma:charactxrows},  $S_{\b}$ satisfies the isometry-dual property if and only if there exists ${\bf x}_{\b}=(x_1,\dots,x_n)$ with non-zero components in the kernel of the matrix with rows in
      $$\left((\varphi_\ii(P_1)\varphi_\jj(P_1),\dots,\varphi_\ii(P_n)\varphi_\jj(P_n))\right)_{\ii+\jj\leq n},$$
      i.e. such that
      $$\sum_{i=1}^n\varphi_\ii(P_i)\varphi_\jj(P_i)x_i=0,$$
      for all $\ii,\jj$ with $\ii+\jj\leq n$,
      i.e. such that
      $$\sum_{i=1}^n(\tau_i \cdot \varphi_\ii(P_i))(\tau_i\cdot  \varphi_\jj(P_i))\frac{x_i}{\tau_i^2}=0,$$
      for all $\ii,\jj$ with $\ii+\jj\leq n$,
      i.e. such that
      $$\sum_{i=1}^n\frac{(f  \varphi_\ii)(P_i)}{f(P_1)}\frac{(f \varphi_\jj)(P_i)}{f(P_1)}\frac{x_i}{\tau_i^2}=0,$$
      for all $\ii,\jj$ with $\ii+\jj\leq n$.
      And this occurs if and only if $S_{(\beta_1+u_1,\dots,\beta_t+u_t)}$ satisfies the isometry-dual property with respect to
$$x_{(\beta_1+u_1,\dots,\beta_t+u_t)}=\left(\frac{x_1}{\tau_1^2}, \frac{x_2}{\tau_2^2},\dots,\frac{x_n}{\tau_n^2}\right).$$
  \end{enumerate}
\end{proof}

We notice that for Kummer extensions of type $y^m= \prod_{i=1}^t (x -\a_i), \a_i \in \fq$ and $(m, t)=1$, the existence of a function as in (\ref{functionf}) in Theorem \ref{t:translationplus} is always guaranteed. In fact, fix $P_\infty$ the only place at infinity, and $Q_1, \dots, Q_t$ the totally ramified places in $\fq(x,y)$ over the places in $\fq(x)$ corresponding to $\a_1, \dots, \a_t$. Let $a \geq 1$ and $b_i \geq a/m$, then we have  the following divisor
$$\left( \frac{y^a}{\prod_{i=1}^t (x-\a_i)^{b_i}} \right)=(m \sum_{i=1}^t b_i-at)P_\infty-\sum_{i=1}^t (mb_i -a)Q_i.$$
In \cite{BCQ2023} the existence of flags satisfying the isometry-dual property for these Kummer extensions has been proved. 

We wonder about the existence of flags satisfying the isometry-dual property in other families of function fields. In the positive case it is also interesting to ask about the existence of a function with divisor as in (\ref{functionf}).

\section{{Construction of self-orthogonal flags of codes}}
\label{s:sufficientcondition}

In recent years, the interest on AG codes with certain special properties has been renewed due to their important role in applications. In particular, self-orthogonal codes can be used to construct quantum error-correcting codes, see \cite{KM2008, MTF2016, BMZ2021}. In this section, we give a sufficient condition for the existence of self-orthogonal flags of AG codes, and propose a method for their construction, should the necessary condition be satisfied.

  \begin{lemma}\label{l:pretselfdualplus}
    Let $\cF$ be a function field of genus $g$ defined over a finite field $\fq$ of characteristic two. Fix  $P, P_1, \ldots, P_n, Q_1, \dots ,Q_t, n \geq 2g+2$ pairwise distinct rational places in $\cF$.
    Suppose that the flag of AG codes $S_{\bf \b}$ as defined in \eqref{defS} satisfies the isometry-dual property with respect to a vector ${\bf x}_\b=(x_1, \ldots, x_n)$ in $(\fq^*)^n$. If there exists a rational function $h$
    such that
    $$(h(P_1),\dots,h(P_n))=\left(\frac{\sqrt{x_1}}{x_1}, \frac{\sqrt{x_1x_2}}{x_1x_2}, \ldots , \frac{\sqrt{x_1x_n}}{x_1x_n}\right),$$
    and such that
    $$(h)=u_1Q_1+\dots +u_tQ_t-uP$$  for some $u, u_1,\dots,u_t \in \N$, then, the flag $S_{(\beta_1+u_1,\dots,\beta_t+u_t)}$ is a self-orthogonal flag.
\end{lemma}

    \begin{proof}
    Consider the function
   $f=\frac{1}{h}$.
    It satisfies $(f)=uP-u_1Q_1-\dots-u_tQ_t$.
    Then, by Theorem~\ref{t:translationplus},
    $S_\b$ satisfies the isometry-dual property with respect to the vector ${\bf x}_\b$ if and only if $S_{(\beta_1+u_1,\dots,\beta_t+u_t)}$ satisfies the isometry-dual property with respect to the vector ${\bf x}_{(\beta_1+u_1,\dots,\beta_t+u_t)}$ defined componentwise by 
$({\bf x}_{(\beta_1+u_1,\dots,\beta_t+u_t)})_i=\frac{({\bf x}_\b)_i}{\tau_i^2}$
    with 
    $$\tau=\left(1,\frac{f(P_2)}{f(P_1)},\dots,\frac{f(P_n)}{f(P_1)}\right)=\left(1,\frac{h(P_1)}{h(P_2)},\dots,\frac{h(P_1)}{h(P_n)}\right).$$

    Hence, 
    $$({\bf x}_{(\beta_1+u_1,\dots,\beta_t+u_t)})_i=\dfrac{(h(P_i))^2}{(h(P_1))^2}({\bf x}_\b)_i=\dfrac{x_1}{x_1x_i}x_i=1.$$
    \end{proof}
   
\begin{lemma}\label{funcaof}
  Let $\cF/\fq$ be a function field.
Fix pairwise distinct rational places $P, \tilde P_1, \ldots, \tilde P_{\tilde n}$ in $\cF$
  and consider the divisor $\tilde D=\sum_{i=1}^{\tilde n} \tilde P_i$.
  Let $1\leq m_1 <m_2 < \dots <m_{\tilde n}$ be the integers such that $\dim C_\cL(\tilde D, m_i P)=i$ and let $f_i$ in $\cL(m_i P)$ be such that the vectors $(f_1(\tilde P_1),\dots,f_1(\tilde P_{\tilde n})), \dots, (f_i(\tilde P_1),\dots,f_i(\tilde P_{\tilde n}))$ form a basis of the code $C_\cL(\tilde D,m_iP)$. Then, for each vector ${\bf y}=(y_1, \ldots, y_{\tilde n})$ in $\fq^{\tilde n}$ there exists a unique linear combination $f_{\bf y}$ of the functions $f_1,\dots,f_{\tilde n}$, such that $(f_{\bf y}(\tilde P_1), \dots , f_{\bf y}(\tilde P_{\tilde n}))=(y_1, \dots , y_{\tilde n})$.
  
\end{lemma}
\begin{proof}
Since $C_\cL(\tilde D, m_{\tilde n} P)=\fq^{\tilde n}$, given $\bf y$, there exists unique $\lambda_1, \dots, \lambda_{\tilde n} $ in $\fq$ such that 
$$(y_1, \ldots, y_{\tilde n})=\lambda_1(f_1(\tilde P_1),\dots,f_1(\tilde P_{\tilde n}))+ \cdots +\lambda_{\tilde n}(f_{\tilde n}(\tilde P_1),\dots,f_{\tilde n}(\tilde P_{\tilde n}))$$

Define $f_{\bf y}=\lambda_1f_1 + \cdots +\lambda_{\tilde n}f_{\tilde n}$.
\end{proof}

The function $f_{\bf y}$ in the previous lemma has only poles  in $P$. 
We also notice that in the previous Lemma \ref{funcaof} a basis for the codes $C_\cL(\tilde D, m_i\tilde P)$ can be chosen taking $\tilde n$ functions $f_i$ in $\cL(m_{\tilde n} P)$ such that $v_{P}(f_i)=-m_i$. Then, for any $i=1, \dots, \tilde n$ we have that the vectors $$v_1=(f_1(\tilde P_1), \dots , f_1(\tilde P_{\tilde n})), \dots , v_i=(f_i(\tilde P_1), \dots f_i(\tilde P_{\tilde n}))$$ form a basis for the code $C_\cL(\tilde D, m_i P)$.

The next theorem is a formalization of the method we propose for finding self-orthogonal flags of codes.

\begin{theorem}\label{t:pretselfdualplus}
Let $\cF$ be a function field of genus $g$ defined over a finite field $\fq$ of characteristic two. Fix  $P, P_1, \ldots, P_n, Q_1, \dots ,Q_t, n \geq 2g+2$ pairwise distinct rational places in $\cF$.
Suppose that the flag of AG codes $S_{\bf \b}$ as defined in \eqref{defS} satisfies the isometry-dual property with respect to a vector ${\bf x}_\b=(x_1, \ldots, x_n)$ in $(\fq^*)^n$. Let 
 $${\bf y}=\left(\frac{\sqrt{x_1}}{x_1}, \frac{\sqrt{x_1x_2}}{x_1x_2}, \ldots , \frac{\sqrt{x_1x_n}}{x_1x_n},\underbrace{0,\dots,0}_{t}\right).$$ 
Let $f_{\bf y}$ be the function given in Lemma \ref{funcaof} with respect to the vector ${\bf y}$ with $\tilde n=n+t$, $\tilde P_1=P_1,\dots,\tilde P_n=P_n,\tilde P_{n+1}=Q_1,\dots,\tilde P_{\tilde n}=Q_t$.
 If $$(f_{\bf  y})=u_1Q_1+\dots +u_tQ_t-uP$$  for some $u, u_1,\dots,u_t \in \N$, then, the flag $S_{(\beta_1+u_1,\dots,\beta_t+u_t)}$ is a self-orthogonal flag. 
\end{theorem}

\begin{proof}
  Lemma~\ref{funcaof} guarantees the existence of a function $f_{\bf y}$
  interpolating ${\bf y}$. Now, if $(f_{\bf  y})=u_1Q_1+\dots +u_tQ_t-uP$ for some $u, u_1,\dots,u_t \in \N$, by Lemma~\ref{l:pretselfdualplus}, the flag $S_{(\beta_1+u_1,\dots,\beta_t+u_t)}$ is a self-orthogonal flag. 
\end{proof}

\begin{example}
\label{ex:hermite}
  Let $\cH$ be the Hermitian function field  over $\mathbb{F}_{16}=\mathbb{F}_2(\alpha)$ with $\alpha^4+\alpha+1=0$, defined by $y^5+x^4+x=0$ of genus $g=6$. We consider two-point codes over ${\mathbb F}_{16}$ with $\b=\beta_1=2$, $P=P_\infty$ the only place at infinity and $Q$ the place with affine coordinates $(0,0)$. By the results in \cite[Theorem VI.3]{BCQ2022}, the flag given by the codes $C_{\mathcal L}(D, aP_\infty+2Q)$ satisfies the isometry-dual property, where $D$ is the sum of all rational places except for $P_\infty$ and $Q$. We considered the sum of points in $D$ in the following order, where we only represented the affine coordinates:
\\\resizebox{\textwidth}{!}{$(\alpha,1)+
(\alpha,\alpha^{3})+
(\alpha,\alpha^{9})+
(\alpha,\alpha^{6})+
(\alpha,\alpha^{12})+
(\alpha^{2},1)+
(\alpha^{2},\alpha^{3})+
(\alpha^{2},\alpha^{9})+
(\alpha^{2},\alpha^{6})+
(\alpha^{2},\alpha^{12})+$}
\resizebox{\textwidth}{!}{$(\alpha^{3},\alpha^{2})+
(\alpha^{3},\alpha^{8})+
(\alpha^{3},\alpha^{5})+
(\alpha^{3},\alpha^{14})+
(\alpha^{3},\alpha^{11})+
(\alpha^{4},1)+(\alpha^{4},\alpha^{3})+
(\alpha^{4},\alpha^{9})+
(\alpha^{4},\alpha^{6})+
(\alpha^{4},\alpha^{12})+$}\\
\resizebox{\textwidth}{!}{$(\alpha^{5},0)+
(\alpha^{6},\alpha)+
(\alpha^{6},\alpha^{4})+
(\alpha^{6},\alpha^{10})+
(\alpha^{6},\alpha^{7})+
(\alpha^{6},\alpha^{13})+
(\alpha^{7},\alpha)+
(\alpha^{7},\alpha^{4})+
(\alpha^{7},\alpha^{10})+
(\alpha^{7},\alpha^{7})+
  (\alpha^{7},\alpha^{13})+$}
\mut{
(\alpha^{8},1)
(\alpha^{8},\alpha^{3})
(\alpha^{8},\alpha^{9})
(\alpha^{8},\alpha^{6})
(\alpha^{8},\alpha^{12})
(\alpha^{9},\alpha)
(\alpha^{9},\alpha^{4})
(\alpha^{9},\alpha^{10})
(\alpha^{9},\alpha^{7})
(\alpha^{9},\alpha^{13})
(\alpha^{10},0)
(\alpha^{11},\alpha^{2})
(\alpha^{11},\alpha^{8})
(\alpha^{11},\alpha^{5})
(\alpha^{11},\alpha^{14})
(\alpha^{11},\alpha^{11})
(\alpha^{12},\alpha^{2})
(\alpha^{12},\alpha^{8})
(\alpha^{12},\alpha^{5})
(\alpha^{12},\alpha^{14})
(\alpha^{12},\alpha^{11})}
\resizebox{\textwidth}{!}{$\dots(\alpha^{13},\alpha)+
(\alpha^{13},\alpha^{4})+
(\alpha^{13},\alpha^{10})+
(\alpha^{13},\alpha^{7})+
(\alpha^{13},\alpha^{13})+
(\alpha^{14},\alpha^{2})+
(\alpha^{14},\alpha^{8})+
(\alpha^{14},\alpha^{5})+
(\alpha^{14},\alpha^{14})+
(\alpha^{14},\alpha^{11})+(1,0)$.}

In this case the isometry vector is   
$${\bf x}=
(1,1,1,1,1,\alpha,\alpha,\alpha,\alpha,\alpha,\alpha^{2},\alpha^{2},\alpha^{2},\alpha^{2},\alpha^{2},\alpha^{3},\alpha^{3},\alpha^{3},\alpha^{3},\alpha^{3},\alpha^{4},\alpha^{5},\alpha^{5},\alpha^{5},$$ $$\alpha^{5},\alpha^{5},\alpha^{6},\alpha^{6},\alpha^{6},\alpha^{6},\alpha^{6},\alpha^{7},\alpha^{7},\alpha^{7},\alpha^{7},\alpha^{7},\alpha^{8},\alpha^{8},\alpha^{8},\alpha^{8},\alpha^{8},\alpha^{9},\alpha^{10},\alpha^{10},\alpha^{10},\alpha^{10},$$ $$\alpha^{10},\alpha^{11},\alpha^{11},\alpha^{11},\alpha^{11},\alpha^{11},\alpha^{12},\alpha^{12},\alpha^{12},\alpha^{12},\alpha^{12},\alpha^{13},\alpha^{13},\alpha^{13},\alpha^{13},\alpha^{13},\alpha^{14}),$$
  and we can take
  $${\bf y}=(1,1,1,1,1,\alpha^{7},\alpha^{7},\alpha^{7},\alpha^{7},\alpha^{7},\alpha^{14},\alpha^{14},\alpha^{14},\alpha^{14},\alpha^{14},\alpha^{6},\alpha^{6},\alpha^{6},\alpha^{6},\alpha^{6},\alpha^{13},$$ $$\alpha^{5},\alpha^{5},\alpha^{5},\alpha^{5},\alpha^{5},\alpha^{12},\alpha^{12},\alpha^{12},\alpha^{12},\alpha^{12},\alpha^{4},\alpha^{4},\alpha^{4},\alpha^{4},\alpha^{4},\alpha^{11},\alpha^{11},\alpha^{11},\alpha^{11},\alpha^{11},$$ $$\alpha^{3},\alpha^{10},\alpha^{10},\alpha^{10},\alpha^{10},\alpha^{10},\alpha^{2},\alpha^{2},\alpha^{2},\alpha^{2},\alpha^{2},\alpha^{9},\alpha^{9},\alpha^{9},\alpha^{9},\alpha^{9},\alpha,\alpha,\alpha,\alpha,\alpha,\alpha^{8},0).$$
The function interpolating ${\bf y}$ is
$f_{{\bf y}}=\alpha^8x^7$. Then, taking
$f=\alpha^7/x^7$
we get that $(f)= 35P_\infty- 35Q$.
Hence, it can be deduced that the family of codes $C_{\mathcal L}(D, aP_\infty+(2+35)Q)=C_{\mathcal L}(D,aP_\infty+37Q)$, with $a$ ranging in $\{min(H_{(37)}^*)-1\}\cup H_{37}^*$,
constitutes a 
self-orthogonal flag by Theorem \ref{t:pretselfdualplus}.
 In this case, $H_{(37)}^*=\{-35, -31, -30\}\cup\{-28,\dots,28\}\cup\{30, 31, 35\}$. Hence, the dual of $C_{\mathcal L}(D,-36P_\infty+37Q)$ is $C_{\mathcal L}(D,35P_\infty+37Q)$, the dual of $C_{\mathcal L}(D,-35P_\infty+37Q)$ is $C_{\mathcal L}(D,31P_\infty+37Q)$, the dual of $C_{\mathcal L}(D,-31P_\infty+37Q)$ is $C_{\mathcal L}(D,30P_\infty+37Q)$, and so on.
\end{example}

\begin{example}
  \label{ex:koetter}
Using the results in Theorem \ref{t:pretselfdualplus} we found some interesting examples of
self-orthogonal complete flags of codes over maximal curves.
For $\ell=2$ and $\ell=3$ we considered the curve $\mathcal{X}_\ell: y^{2^\ell+1}=x^{2^{\ell-1}}+x^{2^{\ell-2}}+\dots+x^2+x$ over ${\mathbb F}_{2^{2\ell}}$, which is maximal (with $2^{3\ell-1}$ points). 
This is a particular example of the maximal curves analyzed by K\"oter in \cite{Koetter1997}. They also appeared in \cite{AT1999} in the context of the classification of maximal curves.
Let $P=P_\infty$ and $Q=(0,0)$. 
  We obtained that
  \begin{itemize}
  \item For $\ell=2$, the flag $C_{\mathcal L}(D, aP_\infty + 37Q)$ is a self-orthogonal flag.
  \item For $\ell=3$, the flag $C_{\mathcal L}(D, aP_\infty + 283Q)$ is a self-orthogonal flag.
  \end{itemize}
  
  We let it as an open question to verify whether it may be true that for $\ell\geq 2$, the sequence $C(D, aP_\infty+(2^{3\ell-1}+2^{2\ell-1}-2^{\ell -1}-1)Q)$ is a self-orthogonal flag.

  In the Appendix we explain how we obtained these results.
\end{example}

\section{Characterizing the divisors giving the isometry-dual property}
\label{s:charact}

In this section we characterize the tuples $\beta$ such that the corresponding flags $S_\beta$ associated to the divisor ${\mathbf G}_\beta$ satisfy the isometry-dual property. We also characterize the corresponding isometry vectors ${\mathbf x}_\beta$.

\begin{definition}
Let $\cF/\fq$ be a function field and let $P, Q$ be different rational places in $\cF$. The period $\pi(P, Q)$ of $P, Q$ is $$\pi(P, Q)= \min \{ u \geq 1 \mid u(P-Q) \text{ is a principal divisor }\}\,.$$
\end{definition}
We notice that the period of two places is always finite since the divisor class group in a function field $\cF$ is finite, see \cite[Proposition 5.1.3]{St2009}. In general, it is a difficult task to determine the period of two rational points. For Kummer extensions with only one place at infinity $P_\infty$ defined by $y^m=f(x)$, where $f(x)$ is a separable polynomial, and $P_1, P_2$ are two totally ramified places distinct from $P_\infty$ the period  $\pi(P_1, P_2)$ is $m$, see  \cite[Proposition 4.9]{CMT}. We now observe that in Kummer extensions, in fact, the period $\pi(P_\infty, P_1)$ is also $m$.
\begin{lemma}\label{l:periodPinf}
Let $y^m=f(x), f(x)$ a separable polynomial of degree $r$, where $ 2 \leq r <m, \gcd(m, r)=1$. Then, the period $\pi(P_\infty, P_\alpha)$ equals $m$, where $P_\alpha$ in the only zero of $x-\alpha$ for $\alpha$ a root of $f(x)$.
\end{lemma}
\begin{proof}
The fact that $(x-\alpha)=m(P_\alpha-P_\infty)$ shows that $\pi\leq m$.
Suppose that $1\leq \pi\leq m-1$ and let $z$ be a rational function such that $(z)=\pi(P_\alpha- P_\infty)$. By \cite[Proposition 4.3]{CMQ2024}, the smallest non-zero element in
the Weierstrass semigroup $H(P_\alpha)$ at $P_\alpha$
is $m-\lfloor m/r \rfloor$. Therefore, $m-\lfloor m/r \rfloor\leq \pi\leq m-1$. Since we have the equality
$$(z(x-\alpha)^{-1})=(m-\pi)(P_\infty-P_\alpha),$$ we obtain that $m-\pi$
is also in $H(P_\alpha)$.
But $0<m-\pi\leq m- (m-\lfloor m/r \rfloor)=\lfloor m/r \rfloor<m-\lfloor m/r \rfloor$, a contradiction. We conclude that $\pi=\pi(P_\infty,P_\alpha)=m$.
\end{proof}

Suppose $\pi =\pi(P, Q)$ and $f \in  \cF$ is such that  $(f)=\pi(P-Q)$. Then, for any rational place $P_1 \notin \{P, Q\}$, we have $f(P_1)=a$ for some $a \in \fq^*$ and, by defining $\varphi = f/a$, we get $\varphi(P_1)=1$. Without loss of generality, we can choose $f$ such that $(f)=\pi(P-Q)$ and $f(P_1)=1$ for a fixed $P_1 \notin \{P, Q\}$.

\begin{definition}\label{defT}
Let $P, Q_1, \dots, Q_t$ be pairwise distinct rational places in $\cF$, and $\pi_i:=\pi(P, Q_i)$ the period of $P$ and $Q_i$. Define
\begin{align*}
&T=\{(\delta_1, \dots, \delta_t) \in \Z^t \mid  \sum_{i=1}^t \delta_i(P-Q_i) \text{ is a principal divisor}\}, \text{ and}\\
&T_0=\{(\delta_1, \dots, \delta_t) \in T  \mid  0 \leq \delta_i < \pi_i \mbox{ for all }i \}.
\end{align*}
\end{definition}

\begin{proposition} \label{TT}
The following statements hold:
\begin{enumerate}[i)]
\item $T$ is an additive subgroup of $\Z^t$.
\item $T_0$ is a group with the operation $\oplus$ defined by:
$$(a_1, \dots, a_t) \oplus (b_1, \dots, b_t) = (a_1+b_1 \mod{\pi_1},\ \ \dots\ \ , a_t+b_t \mod{\pi_t}).$$
\item $T=T_0+ \sum_{i=1}^t \Z \pi_i e_i$, where $e_i$ is the $i$-th canonical vector of ${\mathbb Z}^t$.
\end{enumerate}
\end{proposition}
\begin{proof} The first item follows from the fact that the set of principal divisors in $\cF$ is a group. We now prove the second item. Notice that the zero vector is in $T_0$. Let $(a_1, \dots, a_t)$ and $ (b_1, \dots, b_t) $ be in $T_0$, then there exist functions $\varphi_a$ and $\varphi_b$ such that $(\varphi_a)=\sum_{i=1}^t a_i(P-Q_i), (\varphi_b)=\sum_{i=1}^t b_i(P-Q_i)$. Since $\pi_i=\pi(P, Q_i)$ there exist also functions $\varphi_i$ such that $(\varphi_i)=\pi_i(P-Q_i)$. Let $r_k \equiv a_k+b_k \mod{\pi_k}$, that is $a_k+b_k=\pi_k q_k +r_k, q_k \in \Z,   0 \leq r_k < \pi_k$. Then we have 
  \begin{eqnarray*}
      \sum_{i=1}^t r_i (P-Q_i)&=&
      \sum_{i=1}^t a_i (P-Q_i)+\sum_{i=1}^t b_i (P-Q_i)-\sum_{i=1}^t q_i\pi_i (P-Q_i)\\
      &=&
      (\varphi_a)+(\varphi_b)-\sum_{i=1}^t q_i(\varphi_i)
      \\
      &=&
      (\varphi_a \varphi_b \varphi_1^{-q_1}\cdot \ldots \cdot \varphi_k^{-q_k}).
      \end{eqnarray*}
  We conclude that $(r_1, \dots, r_k) \in T_0$. The inclusion  $T_0+ \sum_{i=1}^t \Z \pi_i e_i \subseteq T$ follows from the facts that $T_0,  \Z \pi_i e_i \subset T$ for $i=1, \dots, t$ and $T$ is a group. For the inclusion $T \subseteq T_0+ \sum_{i=1}^t \Z \pi_i e_i$, given $(\delta_1, \dots, \delta_t) \in T$ it is enough to divide each $\delta_i$ by $\pi_i$ and get a remainder $0\leq r_i< \pi_i$ and make a similar argument as in the proof of the second item.
\end{proof}

\begin{remark}\label{iso-function}
Fix $P, P_1, \dots, P_n, Q_1, \dots, Q_t$ pairwise different rational places of $\cF$.
Let $\b=(\b_1, \dots, \b_t)$ be a $t$-tuple in $\Z^t$ and define ${\bf G}_\b=\sum_{i=1}^t \b_i Q_i$. Define  a second divisor $D=\sum_{i=1}^n P_i$.
Let $W$ be a  Weil differential $W$ such that $v_{P_i}(W)=-1$ and $W_{P_i}(1)=1$ for any $P_i$ in $\supp(D)$ (see \cite[Proposition 2.2.10]{St2009}). In \cite[Theorem 4.2]{BCQ2023}, we proved  that 
$S_{(\b_1,\dots ,\b_t)}$ satisfies the isometry-dual property with respect the isometry vector ${\bf{x}}$ if and only if the divisor 
$$E_\b=(n+2g-2-2\sum_{i=1}^{t} \b_i)P_\infty+2{\bf G_\b}-D$$ 
is canonical,
that is, if and only if the divisor $W-E_\b$ is a principal divisor. In this case, let $f_{\bf \beta}$ be such that $W-E_\b=(f_{\bf \beta})$. The vector ${\bf x}=ev_D(f_{\bf \beta})$ is the isometry vector giving the isometry-dual property to $S_{\bf \beta}$.  
\end{remark}

\begin{theorem}\label{t:T}
Let $\b=(\b_1, \dots, \b_t)$ be a $t$-tuple in $\Z^t$ and let $S_\b$ be a flag of codes in ${\mathbb F}_q^n$ as defined in equation \eqref{defS}. Let $\pi_i= \pi(P, Q_i)$. Suppose $S_\b$ satisfies the isometry-dual property with respect to the isometry vector ${\bf{x}_\b}$.   
\begin{enumerate}[i)]
	\item For $\gg$ in $\Z^t$  the flag $S_\gg$ satisfies the isometry-dual property if and only if 
	\begin{equation}\label{gamma}
	\gg=(\g_1, \dots, \g_t)=\left(\b_1+\frac{\theta_1+ \lambda_1\pi_1}2, \dots, \b_t+\frac{\theta_t+ \lambda_t\pi_t}2\right)
	\end{equation} for some $(\theta_1, \dots, \theta_t)$ in $T_0$, $\lambda_i \in \Z, i=1, \dots, t$ such that $\theta_i+\lambda_i \pi_i$ is even for  $i=1, \dots, t.$
In this case, the isometry vector is given by
\begin{equation}
\label{eqgamma}
      {\bf{x}}_\gg= {\bf{x}}_\b*((\psi_{\bf{\theta}}\prod_{i=1}^t \varphi_i^{\lambda_i})(P_1), \dots , (\psi_{\bf{\theta}}\prod_{i=1}^t\varphi_i^{\lambda_i})(P_n)),
\end{equation}
where the divisor of the function $\psi_{\bf\theta}$ is $(\psi_{\bf\theta})=\sum_{i=1}^t  \theta_i(P-Q_i)$ and $(\varphi_i )=\pi_i(P-Q_i)$ for $i=1, \dots , t.$
	
\item If $\tilde{\lambda}_1,\ldots,\tilde{\lambda}_t$ are such that $\tilde{\lambda}_i\equiv 0\pmod{q-1},$ and
$\tilde{\lambda}_i\pi_i\equiv 0\pmod{2}\,,$ for all $i$,
    then 
	  $${{\bf{x}}_\b}={\bf{x}}_{\b+\left( \frac{\tilde{\lambda}_1\pi_1}{2},\ldots,\frac{\tilde{\lambda}_t\pi_t}{2} \right)}\,$$
        and,
          \begin{itemize}
          \item if $q$ is even,
            then $${S_\b}=S_{\b+\left( \frac{\tilde{\lambda}_1\pi_1}{2},\ldots,\frac{\tilde{\lambda}_t\pi_t}{2} \right)},$$ that is, each code of the flag $S_\b$ equals the corresponding (with same dimension) code of the flag $S_{\b+\left( \frac{\tilde{\lambda}_1\pi_1}{2},\ldots,\frac{\tilde{\lambda}_t\pi_t}{2} \right)}$.

          \item otherwise,
            \begin{equation}\label{eq:u}{S_\b}=v* S_{\b+\left( \frac{\tilde{\lambda}_1\pi_1}{2},\ldots,\frac{\tilde{\lambda}_t\pi_t}{2} \right)}\,\end{equation}

            for some $v\in\{-1,1\}^n$, in the sense that each code of the flag $S_\b$ is $v$-isometric to the corresponding (with same dimension) code of the flag $S_{\b+\left( \frac{\tilde{\lambda}_1\pi_1}{2},\ldots,\frac{\tilde{\lambda}_t\pi_t}{2} \right)}$.
            
            \end{itemize}
          
        \item
Suppose there exists a vector $\beta\in{\mathbb F}_q^t$ such that $S_\beta$ satisfies the isometry-dual property.
Let
\begin{eqnarray*}
e&=&\#\{i: \pi_i\equiv 0\mod 2\},\\
o&=&\#\{i: \pi_i\equiv 1\mod 2\},\\\\
T_0^{(0)}&=&\#\{\theta\in T_0:\not\exists i\mbox{ with }\theta_i\mbox{ odd and }\pi_i\mbox{ even}\},\\
\end{eqnarray*}
Then,
the number of different isometry vectors is at most 
$$\left\{\begin{array}{ll}
T_0^{(0)}(q-1)^t&\mbox{if }q\mbox{ is even,}\\
T_0^{(0)}\left(\frac{q-1}{2}\right)^o(q-1)^e&\mbox{otherwise.}\\
\end{array}
\right.$$
and the number of different isometry-dual flags is at most the number of 
different isometry vectors if $q$ is even or
the number of 
different isometry vectors if $q$ times $2^{t-1}$, otherwise.
\end{enumerate}
\end{theorem}
\begin{proof}
With notation as in Remark \ref{iso-function},
suppose the flag $S_\b$ satisfies the isometry-dual property, then $W-E_\b=W-(n+2g-2-2\sum_{i=1}^t \b_i)P_\infty-2{\bf G_\b}+D$ is a principal divisor. We have that  $S_\gg$ satisfies the  isometry-dual property if and only if 
$W-E_\gg=W-(n+2g-2-2\sum_{i=1}^t \gg_i)P_\infty-2{\bf G_\gg}+D$ is a principal divisor as well. This is equivalent to the divisor 
$$W-E_\gg-(W-E_\b)=2\sum_{i=1}^t (\g_i -\b_i)(P-Q_i)$$ 
being principal, that is, $2(\g_1-\b_1, \dots, \g_t-\b_t)$ belonging to $T=T_0+ \sum_{i=1}^t \Z \pi_i e_i$, from Proposition \ref{TT}. Then, equation \eqref{gamma} follows. 

Let $\psi_{\bf\theta}, \varphi_i , i=1, \dots t$ be functions such that $(\psi_{\bf\theta})=\sum_{i=1}^t  \theta_i(P-Q_i)$ and $(\varphi_i )=\pi_i(P-Q_i)$. We have that ${\bf{x}_\b}=ev_D(f_\b)$, where $(f_\b)=W-E_\b$, by Remark \ref{iso-function}. In the same way, we have that ${\bf{x}_\gg}= ev_D(f_\gg)$ where
\begin{align*}
(f_\gg)&=W-E_\gg\\
&=(f_\b)+2\sum_{i=1}^t (\g_i-\b_i)(P-Q_i)\\
&=(f_\b)+2\sum_{i=1}^t \frac{\theta_i+\lambda_i \pi_i}2(P-Q_i) \quad \text{(from the first part of the proof)}\\
&=(f_\b)+\sum_{i=1}^t  \theta_i(P-Q_i) + \sum_{i=1}^t  \lambda_i \pi_i(P-Q_i)\\
&=(f_\b)+(\psi_{\bf\theta})+(\prod_{i=1}^t \varphi_i^{\lambda_i})\\
&=(f_\b\psi_{\bf\theta}\prod_{i=1}^t \varphi_i^{\lambda_i}).
\end{align*}
Since  ${\bf{x}_\b} \in (\fq^*)^t$, we have that $f_\b(P_1) \neq 0$ and, since we have the divisor equality $(f_\b)=(f_\b/f_\b(P_1))$, we can suppose $f_\b(P_1)=1$. With an analogous argument  we can suppose $f_\gg(P_1)=1, \psi_{\bf\theta}(P_1)=1, \varphi_i(P_1)=1, i=1, \dots t$. We conclude that $f_\gg=f_\b\psi_{\bf\theta}\prod_{i=1}^t \varphi_i^{\lambda_i}.$
This concludes the proof of i).

For the proof of ii), using the same argument as before and the fact that $\tilde\lambda_i=0\mod q-1$,

${\bf{x}}_{\b+\left( \frac{\tilde{\lambda}_1\pi_1}{2},\ldots,\frac{\tilde{\lambda}_k\pi_k}{2} \right)}=ev_D(f_\b\prod_{i=1}^t \varphi_i^{\tilde\lambda_i})={\bf{x}}_\b$.

We recall now that, given a function $f$ in $ \cF$ with divisor $(f)=uP-u_1Q_1-\dots-u_tQ_t$, the set $\{w_1,\dots, w_\ell\}$ is a basis of $\cL(aP+{\bf G}_\b)$ if and only if $\{f w_1,\dots,f w_\ell\}$ is a basis of $\cL((a-u)P+{\bf G}_{(\beta_1+u_1,\dots,\beta_t+u_t)})$.
Consider now the function $\prod_{i=1}^t \varphi_i^{\frac{\tilde\lambda_i}{2}}$. It is well defined as for even characteristic the square root is well defined and, for odd characteristic, $q-1$ is even, and so is $\tilde\lambda_i$ for all $i$, since $\tilde\lambda_i$ is a multiple of $q-1$. It holds that $$\left(\prod_{i=1}^t \varphi_i^{\frac{\tilde\lambda_i}{2}}\right)=\frac{(\sum\tilde\lambda_i\pi_i)}{2}P-\frac{\tilde\lambda_1\pi_1}{2}Q_1-\dots-\frac{\tilde\lambda_t\pi_t}{2}Q_t.$$
Hence, 
the basis functions of the Riemann-Roch spaces defining the codes in the translated flag
$S_{\gg+\left( \frac{\tilde{\lambda}_1\pi_1}{2},\ldots,\frac{\tilde{\lambda}_t\pi_t}{2} \right)}$
are the same functions as the ones defining the codes in $S_{\gg}$, just multiplied, each, by
$\prod_{i=1}^t \varphi_i^{\frac{\tilde\lambda_i}{2}}$.

Now, if $q$ is even,
$\prod_{i=1}^t \varphi_i^{\frac{\tilde\lambda_i}{2}}=\prod_{i=1}^t \left(\sqrt{\varphi_i}\right)^{\tilde\lambda_i}$, which evaluated at any point gives $1$, since $\tilde\lambda_i$ is a multiple of $q-1$.

If $q$ is odd, then $q-1$ is even and $\frac{\tilde\lambda_i}{2}$ is a multiple of $\frac{q-1}{2}$. Then, 
$\prod_{i=1}^t \varphi_i^{\frac{\tilde\lambda_i}{2}}=\prod_{i=1}^t \left(\varphi_i^{\frac{\tilde\lambda_i}{q-1}}\right)^{\frac{q-1}{2}}$, which evaluated at any point gives either $1$ or $-1$.

Hence, if $q$ is even, the codes in each step of both flags are exactly the same, while if $q$ is odd, the codes in
$S_{\b+\left( \frac{\tilde{\lambda}_1\pi_1}{2},\ldots,\frac{\tilde{\lambda}_t\pi_t}{2} \right)}$
are the same as the codes in $S_\b$, except that some coordinates may be multiplied by $-1$. This is the same as saying that there exists $v\in\{-1,1\}^n$ such that 
${S_\b}=v* S_{\b+\left( \frac{\tilde{\lambda}_1\pi_1}{2},\ldots,\frac{\tilde{\lambda}_t\pi_t}{2} \right)}.$

To prove item iii),
we fix a coordinate $i\in\{1,\dots,t\}$. For each element $\theta\in T_0$ we count the number of choices of $\sigma_i+\lambda_i\pi_i$ such that $\sigma_i+\lambda_i\pi_i$ is even (by i))
modulo $\frac{lcm(2,(q-1)\pi_i)}{2}$ (by ii)).
\begin{itemize}
\item
If $\pi_i$ is even,
\begin{itemize}
\item if $\theta_i$ is even, then $\lambda_i$ can be any value in $\{0,1\dots,q-2\}$, hence, there are $q-1$ choices for $\lambda_i$,
\item if $\theta_i$ is odd, there are no choices.
\end{itemize}
So, if $\pi_i$ is even, then there are either no choices for $\lambda_i$ if $\theta_i$ is odd or $q-1$ choices otherwise.
\item
If $\pi_i$ is odd,
\begin{itemize}
\item if $\theta_i$ is even,
\begin{itemize}
\item
if $q$ is even,
then $\lambda_i$ can be any value in
$\{0,2,4,\dots,2(q-1)-2\}$, hence, there are $q-1$ choices for $\lambda_i$,
\item if $q$ is odd,
then $\lambda_i$ can be any value in
$\{0,2,4,\dots,q-3\}$, hence, there are $\frac{q-1}{2}$ choices for $\lambda_i$,
\end{itemize}
\item if $\theta_i$ is odd,
\begin{itemize}
\item
if $q$ is even,
then $\lambda_i$ can be any value in
$\{1,3,5,\dots,2(q-1)-1\}$, hence, there are $q-1$ choices for $\lambda_i$,
\item if $q$ is odd,
then $\lambda_i$ can be any value in
$\{1,3,5,\dots,q-2\}$, hence, there are $\frac{q-1}{2}$ choices for $\lambda_i$,
\end{itemize}
\end{itemize}
So, if $\pi_i$ is odd, then there are $q-1$ choices for $\lambda_i$ if $q$ is even and $\frac{q-1}{2}$ choices otherwise.
\end{itemize}
The result on the count of the number of different isometry-dual flags follows, on one hand, from the fact that in characteristic $2$ the vector $v$ in ii) can only be the constant vector ${\bf 1}$.
On the other hand, given a flag $S_\beta$ and a vector $v\in\{-1,1\}^n$, it holds that $v* S_\beta=(-v)* S_\beta$. Hence, we can fix $v_1$ to $1$ and let the other components vary in $\{-1,1\}$.
\end{proof}

\begin{remark}
  In Example~\ref{ex:hermite} we constructed the self-orthogonal flag $S_{37}$ over a field of even characteristic. In  this case,
by Lemma~\ref{l:periodPinf}, 
  we have that the period $\pi(P_\infty,Q)$ is $5$. Now, we deduce from Theorem~\ref{t:T} $ii)$  that any other flag $S_{37+i\cdot 15\cdot 5}$ for $i\in{\mathbb Z}$ is also self-ortogonal. However, all these flags are indeed the same one. 

  The same happens with the flag $S_{37}$ of Example~\ref{ex:koetter} for $\ell=2$. In this case, by Lemma~\ref{l:periodPinf}
  the period is again $\pi(P_\infty,Q)=5$,
and all the flags $S_{37+i\cdot (16-1)\cdot 5}$ for $i\in{\mathbb Z}$ coincide with $S_{37}$.
Also, the flag $S_{283}$ for $\ell=3$ coincides with all the flags $S_{283+i\cdot (64-1)\cdot 9}$ for $i\in{\mathbb Z}$, because,
by Lemma~\ref{l:periodPinf},
the period $\pi(P_\infty,Q)$ is $9$.

Indeed, all self-orthogonal flags that emanate from Theorem~\ref{t:pretselfdualplus} are defined over fields of even characteristic and in all these cases the translations of flags described in
Theorem~\ref{t:T} $ii)$ result in the same flag.

On the other hand,
consider the curve with affine equation $y^5=x^2+x+1$ over ${\mathbb F}_{11^2}$. It is a Kummer example over a field of odd characteristic. Its genus is $2$.
Let us take ${\mathbb F}_{11^2}={\mathbb F}_{11}(\alpha)$ with $\alpha^2 = 4\alpha + 9$.
The curve has one point $P=P_\infty$ at infinity, two totally ramified points, $Q_1=(\alpha^{40}:0:1)$ and $Q_2=(\alpha^{80}:0:1)$, and further $n=120$ different points $P_1,\dots,P_n$.
If we take $\beta_1=1,\beta_2=1$, \cite[Theorem 5.5.]{BCQ2023} shows that $S_{(\beta_1,\beta_2)}$ satisfies the $x$-isometry dual property. 
By
Lemma~\ref{l:periodPinf}, 
the periods $\pi(P_\infty, Q_1)$ and $\pi(P_\infty, Q_2)$ is $5$.
Now, Theorem~\ref{t:T} guarantees that taking $\tilde\beta_1=\tilde\beta_2=1+\frac{120\cdot 5}{2}$, we obtain another isometry dual flag with the same isometry vector.
In this case the two flags of codes are different. However, they are related as shown in Equation~\eqref{eq:u}. Computations show that in this case the vector $v$ in that equation is $$v=( 1,  1,  1,  1,  1, -1, -1, -1, -1, -1,  1,  1,  1,  1,  1,  1,  1,  1,  1,  1,  1,  1,  1,  1,  1,$$ $$-1, -1, -1, -1, -1, -1, -1, -1, -1, -1, 1,  1,  1,  1,  1,  1,  1,  1,  1,  1,  1,  1,  1,  1,  1,  1,  1,  1,  1,  1,$$ $$-1, -1, -1, -1, -1,  1,  1,  1,  1,  1,  1,  1,  1,  1,  1,  1,  1,  1,  1,  1,  1,  1,  1,  1,  1,  1,  1,  1,  1,  1,$$ $$-1, -1, -1, -1, -1,  1,  1,  1,  1,  1,  1,  1,  1,  1,  1,$$ $$-1, -1, -1, -1, -1, -1, -1, -1, -1, -1, -1, -1, -1, -1, -1,  1,  1,  1,  1,  1).$$
\end{remark}

\section{Acknowledgments}

The third author would like to thank the fellowship from CIMPA-ICTP that allowed to visit  Bras-Amorós in Catalonia, and they both thank the Col$\cdot$legi Oficial de Doctors i Llicenciats en Filosofia i Lletres i en Ciències de Catalunya.
The three authors also thank CIRM for the scientific visit in December 2023 and are grateful for the many delicious {\it Île flottant} from Lorence.

The first author was supported by the Ministerio de Ciencia e Innovación, under grant PID2021-124928NB-I00, by the AGAUR under grant 2021 SGR 00115, and by the Project HERMES funded by INCIBE and the European Union NextGenerationEU/PRTR.
The second author was supported by Fundação de Amparo à Pesquisa do Estado de Minas Gerais - Brasil (FAPEMIG) APQ 00696-18 and RED 0013-21.
The third author was supported by Coordenação de Aperfeiçoamento de Pessoal de Nível Superior - Brasil (CAPES) - 001, and by Conselho
Nacional de Desenvolvimento Científico e Tecnológico CNPq 307261/2023-9 Bolsa de produtividade.


        \section*{Appendix}
        In Example~\ref{ex:koetter} we considered the curves
        $\mathcal{X}_\ell: y^{2^\ell+1}=x^{2^{\ell-1}}+x^{2^{\ell-2}}+\dots+x^2+x$ over ${\mathbb F}_{2^{2\ell}}$, which are maximal (with $2^{3\ell-1}+1$ rational points),  with genus $g=2^{\ell-1}(2^{\ell-1}-1)$. 
        They are, in fact, examples of  Kummer extensions with
        $m=2^\ell+1$ and $r=2^{\ell-1}$. We focused on the cases $\ell=2$ and $\ell=3$ \cite{Koetter1997,AT1999}.       
Let $P=P_\infty$ and $Q=(0,0)$.
We consider the two-point (i.e. $t=1$) flags of codes $S_{\beta_1}$ given by codes of the form $C_{\mathcal L}(D, aP_\infty + \beta_1Q)$.
In \cite[Corollary VI.4]{BCQ2022} we proved that if
$m$ is odd and $0\leq \beta_1 \leq n/2-g-1$, the flag $S_{\beta_1}$
satisfies the isometry-dual property, 
if and only if $\beta_1=\frac{m\cdot i-1}{2}$ for some $i$, which is necessarily odd, i.e., $i=2j+1$ for some integer $j$. Hence,
the flag $S_{\beta_1}$
satisfies the isometry-dual property, if and only if
$\beta_1=mj+\frac{m-1}{2}$ for some $0\leq j\leq \lfloor\frac{n-m-2g-1}{2m}\rfloor$. 
For this reason, we took $j=0$ and considered the flags $S_{\beta_1}$ with $\beta_1=\frac{m-1}{2}=2^{\ell-1}.$

  \subsubsection*{Curve $y^5=x^2+x$ ($\ell=2$)}

  In this case,
${\mathbb F}_{16}={\mathbb F}_2(\alpha)$ with $\alpha^4=\alpha+1$,
$n= 31$,
$g= 2$.
  We considered
  $D=P_1+\dots+P_{31}=(\alpha,\alpha)+(\alpha,\alpha^{4})+(\alpha,\alpha^{10})+(\alpha,\alpha^{7})+(\alpha,\alpha^{13})+(\alpha^{2},\alpha^{2})+(\alpha^{2},\alpha^{8})+(\alpha^{2},\alpha^{5})+(\alpha^{2},\alpha^{14})+(\alpha^{2},\alpha^{11})+(\alpha^{4},\alpha)+(\alpha^{4},\alpha^{4})+(\alpha^{4},\alpha^{10})+(\alpha^{4},\alpha^{7})+(\alpha^{4},\alpha^{13})+(\alpha^{5},1)+(\alpha^{5},\alpha^{3})+(\alpha^{5},\alpha^{9})+(\alpha^{5},\alpha^{6})+(\alpha^{5},\alpha^{12})+(\alpha^{8},\alpha^{2})+(\alpha^{8},\alpha^{8})+(\alpha^{8},\alpha^{5})+(\alpha^{8},\alpha^{14})+(\alpha^{8},\alpha^{11})+(\alpha^{10},1)+(\alpha^{10},\alpha^{3})+(\alpha^{10},\alpha^{9})+(\alpha^{10},\alpha^{6})+(\alpha^{10},\alpha^{12})+(1,0)$, and, as explained above, we took $\beta_1=2$.
In this case
$H_b=\{0,\dots,33\}$, while $H_b^*=\{0,\dots,30\}$.
The flag of codes $C_{\mathcal L}(D,aP_\infty+2Q)$ satisfies the isometry-dual property with corresponding isometry vector
$${\bf x}=(1,1,1,1,1,\alpha,\alpha,\alpha,\alpha,\alpha,\alpha^{3},\alpha^{3},\alpha^{3},\alpha^{3},\alpha^{3},\alpha^{4},\alpha^{4},\alpha^{4},\alpha^{4},\alpha^{4},$$ $$\alpha^{7},\alpha^{7},\alpha^{7},\alpha^{7},\alpha^{7},\alpha^{9},\alpha^{9},\alpha^{9},\alpha^{9},\alpha^{9},\alpha^{14}).$$
Hence, the vector ${\bf y}$ of Theorem~\ref{t:pretselfdualplus} is $${\bf y}= (1,1,1,1,1,\alpha^{7},\alpha^{7},\alpha^{7},\alpha^{7},\alpha^{7},\alpha^{6},\alpha^{6},\alpha^{6},\alpha^{6},\alpha^{6},\alpha^{13},\alpha^{13},\alpha^{13},\alpha^{13},\alpha^{13},$$ $$\alpha^{4},\alpha^{4},\alpha^{4},\alpha^{4},\alpha^{4},\alpha^{3},\alpha^{3},\alpha^{3},\alpha^{3},\alpha^{3},\alpha^{8},0).$$
and the function interpolating ${\bf y}$ is
$$f_{\bf y}= \alpha^8x^7.$$
Then, we take
$f= 1/f_{\bf y}=\alpha^7/x^7$
and compute its divisor 
$(f)=35 P_\infty-35Q$.
We deduce that the value $u$ in Theorem~\ref{t:pretselfdualplus} is
$u= 35$.
We conclude that the flag $S_{2+35}=S_{37}$ given by the codes $C_{\mathcal L}(D,aP_\infty+37Q)$ with $a$ ranging in $\{min(H_{(37)}^*)-1\}\cup H_{(37)}^*=\{-36,-35,\dots,-5\}$ is a self-orthogonal flag of codes.

\subsubsection*{Curve $y^9=x^4+x^2+x$ ($\ell=3$)}

  In this case,
${\mathbb F}_{64}={\mathbb F}_2(\alpha)$ with $\alpha^6=\alpha^4 + \alpha^3 + \alpha + 1$,
$n= 255$, $g= 12$.
  We do not copy the divisor $D$ for the sake of brevity.
We chose $\beta_1=2^{\ell-1}=4$, as explained above. In this case,
$H_b=\{0,4,8,9,11,12,13\}\cup\{15,\dots,275\}$ and $H_b^*=\{0,4,8,9,11,12,13\}\cup\{15,\dots,255\}\cup\{257, 258, 259, 261, 262, 266, 270\}$.
The flag of codes $C_{\mathcal L}(D,aP_\infty+4Q)$ satisfies the isometry-dual property with corresponding isometry vector $$x=(1,1,1,1,1,1,1,1,1,\alpha,\alpha,\alpha,\alpha,\alpha,\alpha,\alpha,\alpha,\alpha,\alpha^{3},\alpha^{3},\alpha^{3},\alpha^{3},\alpha^{3},\alpha^{3},\alpha^{3},\alpha^{3},\alpha^{3},$$ $$\alpha^{4},\alpha^{4},\alpha^{4},\alpha^{4},\alpha^{4},\alpha^{4},\alpha^{4},\alpha^{4},\alpha^{4},\alpha^{6},\alpha^{6},\alpha^{6},\alpha^{6},\alpha^{6},\alpha^{6},\alpha^{6},\alpha^{6},\alpha^{6},$$ $$\alpha^{7},\alpha^{7},\alpha^{7},\alpha^{7},\alpha^{7},\alpha^{7},\alpha^{7},\alpha^{7},\alpha^{7},\alpha^{8},\alpha^{9},\alpha^{9},\alpha^{9},\alpha^{9},\alpha^{9},\alpha^{9},\alpha^{9},\alpha^{9},\alpha^{9},$$ $$\alpha^{13},\alpha^{13},\alpha^{13},\alpha^{13},\alpha^{13},\alpha^{13},\alpha^{13},\alpha^{13},\alpha^{13},\alpha^{14},\alpha^{14},\alpha^{14},\alpha^{14},\alpha^{14},\alpha^{14},\alpha^{14},\alpha^{14},\alpha^{14},$$ $$\alpha^{15},\alpha^{15},\alpha^{15},\alpha^{15},\alpha^{15},\alpha^{15},\alpha^{15},\alpha^{15},\alpha^{15},\alpha^{16},\alpha^{16},\alpha^{16},\alpha^{16},\alpha^{16},\alpha^{16},\alpha^{16},\alpha^{16},\alpha^{16},$$ $$\alpha^{17},\alpha^{19},\alpha^{19},\alpha^{19},\alpha^{19},\alpha^{19},\alpha^{19},\alpha^{19},\alpha^{19},\alpha^{19},\alpha^{26},\alpha^{26},\alpha^{26},\alpha^{26},\alpha^{26},\alpha^{26},\alpha^{26},\alpha^{26},\alpha^{26},$$ $$\alpha^{27},\alpha^{27},\alpha^{27},\alpha^{27},\alpha^{27},\alpha^{27},\alpha^{27},\alpha^{27},\alpha^{27},\alpha^{29},\alpha^{29},\alpha^{29},\alpha^{29},\alpha^{29},\alpha^{29},\alpha^{29},\alpha^{29},\alpha^{29},$$ $$\alpha^{31},\alpha^{31},\alpha^{31},\alpha^{31},\alpha^{31},\alpha^{31},\alpha^{31},\alpha^{31},\alpha^{31},\alpha^{33},\alpha^{33},\alpha^{33},\alpha^{33},\alpha^{33},\alpha^{33},\alpha^{33},\alpha^{33},\alpha^{33},$$ $$\alpha^{34},\alpha^{34},\alpha^{34},\alpha^{34},\alpha^{34},\alpha^{34},\alpha^{34},\alpha^{34},\alpha^{34},\alpha^{35},\alpha^{38},\alpha^{38},\alpha^{38},\alpha^{38},\alpha^{38},\alpha^{38},\alpha^{38},\alpha^{38},\alpha^{38},$$ $$\alpha^{39},\alpha^{39},\alpha^{39},\alpha^{39},\alpha^{39},\alpha^{39},\alpha^{39},\alpha^{39},\alpha^{39},\alpha^{44},\alpha^{44},\alpha^{44},\alpha^{44},\alpha^{44},\alpha^{44},\alpha^{44},\alpha^{44},\alpha^{44},$$ $$\alpha^{48},\alpha^{48},\alpha^{48},\alpha^{48},\alpha^{48},\alpha^{48},\alpha^{48},\alpha^{48},\alpha^{48},\alpha^{50},\alpha^{50},\alpha^{50},\alpha^{50},\alpha^{50},\alpha^{50},\alpha^{50},\alpha^{50},\alpha^{50},$$ $$\alpha^{53},\alpha^{53},\alpha^{53},\alpha^{53},\alpha^{53},\alpha^{53},\alpha^{53},\alpha^{53},\alpha^{53},\alpha^{55},\alpha^{55},\alpha^{55},\alpha^{55},\alpha^{55},\alpha^{55},\alpha^{55},\alpha^{55},\alpha^{55},$$ $$\alpha^{56},\alpha^{56},\alpha^{56},\alpha^{56},\alpha^{56},\alpha^{56},\alpha^{56},\alpha^{56},\alpha^{56},\alpha^{59},\alpha^{59},\alpha^{59},\alpha^{59},\alpha^{59},\alpha^{59},\alpha^{59},\alpha^{59},\alpha^{59},$$ $$\alpha^{62},\alpha^{62},\alpha^{62},\alpha^{62},\alpha^{62},\alpha^{62},\alpha^{62},\alpha^{62},\alpha^{62})$$
Hence, the vector ${\bf y}$ of Theorem~\ref{t:pretselfdualplus} is 
$${\bf y}=(1,1,1,1,1,1,1,1,1,\alpha^{31},\alpha^{31},\alpha^{31},\alpha^{31},\alpha^{31},\alpha^{31},\alpha^{31},\alpha^{31},\alpha^{31},$$ $$\alpha^{30},\alpha^{30},\alpha^{30},\alpha^{30},\alpha^{30},\alpha^{30},\alpha^{30},\alpha^{30},\alpha^{30},\alpha^{61},\alpha^{61},\alpha^{61},\alpha^{61},\alpha^{61},\alpha^{61},\alpha^{61},\alpha^{61},\alpha^{61},$$ $$\alpha^{60},\alpha^{60},\alpha^{60},\alpha^{60},\alpha^{60},\alpha^{60},\alpha^{60},\alpha^{60},\alpha^{60},\alpha^{28},\alpha^{28},\alpha^{28},\alpha^{28},\alpha^{28},\alpha^{28},\alpha^{28},\alpha^{28},\alpha^{28},$$ $$\alpha^{59},\alpha^{27},\alpha^{27},\alpha^{27},\alpha^{27},\alpha^{27},\alpha^{27},\alpha^{27},\alpha^{27},\alpha^{27},\alpha^{25},\alpha^{25},\alpha^{25},\alpha^{25},\alpha^{25},\alpha^{25},\alpha^{25},\alpha^{25},\alpha^{25},$$ $$\alpha^{56},\alpha^{56},\alpha^{56},\alpha^{56},\alpha^{56},\alpha^{56},\alpha^{56},\alpha^{56},\alpha^{56},\alpha^{24},\alpha^{24},\alpha^{24},\alpha^{24},\alpha^{24},\alpha^{24},\alpha^{24},\alpha^{24},\alpha^{24},$$ $$\alpha^{55},\alpha^{55},\alpha^{55},\alpha^{55},\alpha^{55},\alpha^{55},\alpha^{55},\alpha^{55},\alpha^{55},\alpha^{23},\alpha^{22},\alpha^{22},\alpha^{22},\alpha^{22},\alpha^{22},\alpha^{22},\alpha^{22},\alpha^{22},\alpha^{22},$$ $$\alpha^{50},\alpha^{50},\alpha^{50},\alpha^{50},\alpha^{50},\alpha^{50},\alpha^{50},\alpha^{50},\alpha^{50},\alpha^{18},\alpha^{18},\alpha^{18},\alpha^{18},\alpha^{18},\alpha^{18},\alpha^{18},\alpha^{18},\alpha^{18},$$ $$\alpha^{17},\alpha^{17},\alpha^{17},\alpha^{17},\alpha^{17},\alpha^{17},\alpha^{17},\alpha^{17},\alpha^{17},\alpha^{16},\alpha^{16},\alpha^{16},\alpha^{16},\alpha^{16},\alpha^{16},\alpha^{16},\alpha^{16},\alpha^{16},$$
$$\dots$$
$$\alpha^{2},\alpha^{2},\alpha^{2},\alpha^{2},\alpha^{2},\alpha^{2},\alpha^{2},\alpha^{2},\alpha^{2},\alpha^{32},\alpha^{32},\alpha^{32},\alpha^{32},\alpha^{32},\alpha^{32},\alpha^{32},\alpha^{32},\alpha^{32},0)$$
and the function interpolating ${\bf y}$ is
$f_{\bf y}=\alpha^{32}x^{31}$, 
Then, we take
$f= 1/f_{\bf y}=\alpha^{31}/x^{31}$
and compute its divisor 
$(f)= 279 P_\infty- 279Q$.
We deduce that the value $u$ in Theorem~\ref{t:pretselfdualplus} is
$u=279$.
We conclude that the flag $S_{4+279}=S_{283}$ given by the codes $C_{\mathcal L}(D,aP_\infty+283Q)$ with $a$ ranging in 
  $\{min(H_{(283)}^*)-1\}\cup H_{(283)}^*$
  is a self-orthogonal flag of codes. In this case
  {\small
  \begin{eqnarray*}\{min(H_{(283)}^*)-1\}\cup H_{(283)}^*&=&\{-280,-279,-275,-271,-270,-268,-267,-266\}\\&&\cup\{-264,\dots,-24\}\\&&\cup\{-22, -21, -20, -18, -17, -13, -9\}\;.\end{eqnarray*}}

\end{document}